\documentclass[aps,prd,showpacs,nofootinbib,twocolumn]{revtex4}
\usepackage{latexsym}
\usepackage{amsmath,amsfonts}
\usepackage{amsbsy}
\usepackage{mathrsfs}
\usepackage{color}
\usepackage{float}
\usepackage{dsfont}
\usepackage{psfrag}
\usepackage{enumerate}
\usepackage{soul}
\usepackage{gensymb}
\usepackage{bm}

\usepackage{hyperref}% add hypertext capabilities
\usepackage{color}
\usepackage{ulem}

\newcommand{\f}[2]{\frac{#1}{#2}}

\usepackage{amsmath,amssymb,calc,amsfonts}
\usepackage{latexsym}
\usepackage{hyperref}

\usepackage{graphicx,calc,epsfig}

\begin{document}

\title{
 Higher order corrections to deflection angle of massive particles and light rays in plasma media for stationary spacetimes using the Gauss-Bonnet theorem
}
 
\date{\today}

 \author{Gabriel Crisnejo$^{1,2}$, Emanuel Gallo${^{1,3}}$ and Kimet Jusufi$^{4}$} 

 \affiliation{ 
 $^1$FaMAF, UNC; Ciudad Universitaria, (5000) C\'ordoba, Argentina. \\ 
 $^2$Centro Brasileiro de Pesquisas F\'isicas, Rio de Janeiro, RJ 22290-180, Brasil. \\
 $^3$Instituto de F\'isica Enrique Gaviola (IFEG), CONICET, \\
 Ciudad Universitaria, (5000) C\'ordoba, Argentina.\\
 $^4$Physics Department, State University of Tetovo, Ilinden Street nn, 1200,
Tetovo, North Macedonia}

\begin{abstract}
The purpose of this article is twofold. First,  we extend the results presented in \cite{Crisnejo-gauss-bonnet-1} to stationary spacetimes. Specifically, we show that the Gauss-Bonnet theorem can be applied to describe the deflection angle of light rays in plasma media in stationary spacetimes. Second, by using a correspondence between the motion of light rays in a cold non magnetized plasma and relativistic test massive particles we show that this technique is not only powerful to obtain the leading order behavior of the deflection angle of massive/massless particles in the weak field regime but also to obtain higher order corrections. We particularize it to a Kerr background where we compute the deflection angle for test massive particles and light rays propagating in a non homogeneous cold plasma by including third order corrections in the mass and spin parameters of the black hole. 
\end{abstract}

%\pacs{95.30.Sf}
\maketitle

{\sf \footnotesize \scriptsize
%\tableofcontents
}

\vspace{5mm}

\section{Introduction}

General relativity is a geometric theory of gravitation which predicts the existence of black holes (BHs). BHs are extremely important astrophysical objects characterised by many astrophysical processes. For example, they are important to explain the formation of extragalactic jets from a black hole accretion disk and tidal disruption of stars by supermassive black holes, to study the deflection of light and particles, but also to explore various predictions of quantum field theory and quantum gravity through Hawking radiation. The recent announcement of the first image concerning the detection of an event horizon of a supermassive black hole at the center of the neighboring elliptical M87 galaxy by the Event Horizon Telescope (EHT) collaboration is a great triumph of general relativity \cite{Akiyama:2019cqa,Akiyama:2019brx,Akiyama:2019sww,Akiyama:2019bqs,Akiyama:2019fyp,Akiyama:2019eap}. In general, the image of a black hole with a surrounding accretion disk appears distorted due to the strong gravitational lensing effect. In this way, black holes are expected to cast shadows on the bright background which is related to the existence of an event horizon and therefore an unstable photon region \cite{Perlick-photon-region-1}. The shadow image is of great scientific significance, it can help us to probe the geometrical structure of the event horizon and perhaps to measure the angular momentum parameter of the black hole.

The gravitational deflection of light and particles is not only an important tool in astrophysics to test the existence of BHs, or to distinguish BHs from other compact objects but also to characterize the matter distribution at galactic or extragalactic scales.   Quite amazingly, it turns out that the  deflection angle can be linked with the topology of the spacetime by means of the Gauss-Bonnet theorem (GBT) applied to an associated optical geometry. Namely, Gibbons and Werner \cite{Gibbons-gauss-bonnet} showed that, in asymptotically flat and static spacetimes the deflection angle can be obtained by integrating the Gaussian optical curvature over an infinite domain of integration outside the massive body. This result was generalized to stationary spacetimes by Werner applying the Finsler-Randers geometry \cite{Werner:2012rc}. On the other hand, another method was applied by Ishihara et. al. \cite{Ishihara:2016vdc} in order to calculate the deflection angle using the Gauss-Bonnet theorem.  This method was generalized to stationary spacetimes by Ono et.al. \cite{Ono:2017pie,Ono:2018ybw}. 

Recently a new step forward was put by Crisnejo and Gallo \cite{Crisnejo-gauss-bonnet-1} extending the use of the Gibbons-Werner method to more general situations where light rays propagate in a plasma environment. In that work the deflection angle of test massive particles is also studied using the Gauss-Bonnet theorem by taking into account a known correspondence between the
dynamic of light rays in an homogeneous cold nonmagnetized plasma and test massive particles following geodesics at the same spacetime \cite{Kulsrud-1992}. This correspondence was extensively used by other authors in the past. In \cite{BisnovatyiKogan:2010ar} Bisnovatyi-Kogan and Tsupko used this correspondence to calculate the deflection angle of massive particles in static spacetimes in the weak-field approximation while in \cite{Tsupko:2013cqa} and \cite{Tsupko-strong} the correspondence was used to study the strong deflection limit.

In \cite{Crisnejo-charged-particles} it was established for the first time a correspondence between the motion of light rays in a particular non-homogeneous plasma and the one of relativistic test charged massive particles in vacuum. In the same work, the Gauss-Bonnet method was applied to obtain an expression for the deflection angle in terms of the components of the energy-momentum tensor in a plasma environment generalizing in this way previous works restricted to the pure gravity case \cite{Gallo-lens-2011,Gallo-peculiar,Boero:2016nrd,Crisnejo-lens-2018}.

In many astrophysical situations the effect of the plasma medium on the light rays can be neglected but in others, as for example in the case of light rays propagating in the radio frequency band  this is not suitable. A well know example is the influence of the solar corona on light rays propagating near to the Sun. The first works in this regard were made by Muhleman et. al. \cite{Solar-Radio,Muhleman-1970} who studied the influence of the solar corona on the time delay and in the light deflection in the weak field approximation. This is a case where a non-magnetized plasma is a good model of the medium but in other scenarios could be necessary to consider the presence of a magnetic field, that is, to consider a magnetized plasma. A rigorous derivation of the relation dispersion in this kind of plasma was presented by Breuer and Ehlers \cite{Breuer-1980, Breuer-1981a, Breuer-1981b}.
In this work we will only consider cold non-magnetized plasma for which a very detailed treatment was done by Perlick in \cite{Perlick-book}. In the same reference, Perlick derived an exact closed formula for the deflection angle of light
rays in a Kerr metric surrounded by an arbitrary non-homogeneous plasma. That expression is found in integral form in terms of the closest
approach distance and is not limited to weak field regime. 

Due to the importance of the plasma on the light rays its study becomes an active research area in the last years \cite{BisnovatyiKogan:2008yg,
BisnovatyiKogan:2010ar,
Tsupko:2013cqa,Tsupko:2014sca,
Tsupko:2014lta,Perlick:2015vta,
Bisnovatyi-Kogan:2015dxa,
Perlick:2017fio,
Bisnovatyi-Kogan:2017kii, Rogers:2015dla,
Rogers:2016xcc,
Rogers:2017ofq,
2018MNRAS.475..867E, 
Er:2013efa,Yan:2019etp}.

The purpose of this paper is to extend the results presented in \cite{Crisnejo-gauss-bonnet-1} to stationary spacetimes. To do so we will make use of a Finsler-Randers type  metric.
Although many  papers in the literature applies the GBT to different stationary metrics, the problem of higher order correction terms in the deflection angle in stationary spacetimes is not in general discussed. Here we fill this gap by computing in a successfully way the deflection angle in the weak field regime including up to third order correction. 

The plan of the paper is as follows: In Sec. \ref{sec-2}, we shall use the dynamic of light rays in a plasma medium to introduce the Finsler-Randers metric. In Sec. \ref{sec-3}, we review the GBT in stationary spacetimes. In Sec. \ref{sec-4}, we obtain the deflection angle in stationary spacetimes surrounded by a plasma medium at linear order using the approach proposed by Ono, Ishihara and Asada in \cite{Ono:2017pie}.  In Sec. \ref{sec-5} we extend the Werner's method to  calculate the deflection angle of massive particles. In Sec. \ref{sec-6} we study the problem of higher order corrections to the deflection angle by studying the deflection angle of massive particles and also light rays propagating in a nonhomogeneous plasma medium in a Kerr background. Finally in Sec. \ref{sec-7}  we comment on our results. Two appendices with relevant calculations are also included.

 In the following we shall use natural units $G=c=\hbar=1$ and we adopt the index conventions that Greek indices run from 0 to 3, early Latin indices (a,b,c,...) run from 1 to 3 and middle Latin indices (i,j,k,...) run from 1 to 2.
 
\section{Optical metric for stationary axisymmetric spacetimes in a plasma environment}\label{sec-2}

Let us consider a stationary axisymmetric spacetime $(\mathcal{M},g_{\alpha\beta})$ filled with a cold non-magnetized plasma described by the refractive index $n$,
\begin{equation}\label{refra-index}
    n^2(x,\omega(x))=1-\frac{\omega_e^2(x)}{\omega^2(x)},
\end{equation}
where $\omega(x)$ is the photon frequency measured by an observer following the integral curves of a timelike Killing vector field, and $\omega_e(x)$ is the plasma frequency given by,
\begin{equation}\label{K_e}
    \omega_e^2(x)=\frac{4\pi e^2}{m_e} N(x)= K_e N(x),
\end{equation}
where $e$ and $m_e$ are the charge and the mass of the electron, respectively while $N(x)$ is the number density of electrons in the plasma. Following the notation introduced in \cite{BisnovatyiKogan:2008yg} and since then used in other works by other authors we have used the definition,
\begin{equation}
    K_e = \frac{4\pi e^2}{m_e},
\end{equation}
in Eqn. \eqref{K_e}.

Note that only light rays  with $\omega^2(x)>\omega^2_e(x)$ propagate through the plasma.

The dynamic of the light rays in this kind of medium is usually described through the Hamilton's equations associated with the Hamiltonian,
\begin{equation}\label{ham-1}
    \mathcal{H}(x,p)=\frac{1}{2}(g^{\alpha\beta}p_{\alpha\beta}+\omega_e^2(x)),
\end{equation}
with the dispersion relation,
\begin{equation}\label{dispersion}
    \mathcal{H}(x,p)=0.
\end{equation}
As explained by Perlick in \cite{Perlick-book}, this Hamiltonian define a ray-optical structure on $\mathcal{M}$ and even more, under appropriated conditions it is possible to build a reduced ray-optical structure on a 3-dimensional manifold $\hat{\mathcal{M}}$ in order to discuss the spatial paths of  light rays. 

In this section we will carry out the reduction process described in \cite{Perlick-book} (see Theorem 6.5.1 of that reference for a complete description of this process) to build this reduced ray-optical structure for an arbitrary stationary and axisymmetric spacetime given by the line element, 
\begin{equation}
  ds^2=g_{00}dt^2+2g_{03}dtd\varphi+g_{ab}dx^adx^b.
\end{equation}
To do that, we have to restrict our attention to the region on the spacetime where the Killing vector field $\frac{\partial}{\partial t}$ is timelike and the propagation of light rays is allowed, i.e $\omega^2(x)>\omega_e^2(x)$.

First of all we want to remark that the Hamiltonian \eqref{ham-1} can be rewritten as follows,
\begin{equation}\label{ham-2}
    \mathcal{H}=\frac{1}{2}(g^{ij}p_i p_j +g^{33}(p_3+\frac{g^{03}}{g^{33}}p_0)^2-p_0^2 \Omega^2),
\end{equation}
where
\begin{equation}
    \Omega^2:=\frac{(g^{03})^2-g^{00}g^{33}}{g^{33}}-\frac{\omega_e^2}{p_0^2}.
\end{equation}
As the metric is stationary, $p_0$ is a constant of motion which can be identified with the frequency measured by a stationary observer at infinity, that is, $p_0:=-\omega_\infty$.

Because of the basic equation from which the dynamic of light rays can be deduced is the dispersion relation \eqref{dispersion}, we can multiply the Hamiltonian \eqref{ham-2} by a non-zero function without altering the dynamic of the light rays\footnote{This induces a reparametrization in the Hamilton's equations.}. Then, we have
\begin{equation}
    \mathcal{H}=\Omega^2 \tilde{\mathcal{H}},
\end{equation}
where
\begin{equation}\label{hat-hamil-0}
    \tilde{\mathcal{H}}(x^\alpha,p_\alpha)=\frac{1}{2}\bigg\{{\Omega^{-2}}\bigg[g^{ij}p_i p_j +g^{33}(p_3+\frac{g^{03}}{g^{33}}p_0)^2\bigg]-p_0^2\bigg\},
\end{equation}
and as we will see $\Omega$ is a non-zero function within the region of interest. In particular, by making use of the following identities,
\begin{eqnarray}
    \frac{(g^{03})^2-g^{00}g^{33}}{g^{33}}&=&-\frac{1}{g_{00}},\\
    -\frac{g^{03}}{g^{33}}&=&\frac{g_{03}}{g_{00}},
\end{eqnarray}
we can express the Hamiltonian \eqref{hat-hamil-0} as follows,
\begin{equation}
    \tilde{\mathcal{H}}=\frac{1}{2}\bigg(\frac{g^{ij}p_i p_j +g^{33}(p_3-\frac{g_{03}}{g_{00}}p_0)^2}{\Omega^2}-p_0^2\bigg),
\end{equation}
with
\begin{equation}
    \Omega^2=-\frac{1}{g_{00}} \left(1-\frac{\omega_e^2(x)}{(p_0/\sqrt{-g_{00}})^2} \right).
    %\Omega^2=-\frac{n^2}{g_{00}},
\end{equation}

As discussed in \cite{Perlick-book} a reduced ray-optical structure given by the Hamiltonian $\tilde{\mathcal{H}}$ can be simply obtained by replacing the conserved momentum coordinate $p_0$ by the constant $-\omega_\infty$, i.e. the photon frequency measured by an asymptotic observer. Then, the 3-dimensional reduced ray-optical structure is defined by the Hamiltonian $\hat{\mathcal{H}}$ ,
\begin{equation}\label{hamil-reduce}
\begin{split}
    \hat{\mathcal{H}}(x^a,p_a)=&\tilde{\mathcal{H}}(x^a,p_a,p_0=-\omega_\infty)\\
    =&\frac{1}{2}\bigg(\frac{g^{ij}p_i p_j +g^{33}(p_3+\frac{g_{03}}{g_{00}}\omega_\infty)^2}{\Omega^2}-\omega_\infty^2\bigg),
    \end{split}
\end{equation}
where
\begin{equation}
    \Omega^2=-\frac{1}{g_{00}} \left(1-\frac{\omega_e^2(x)}{(\omega_\infty/\sqrt{-g_{00}})^2} \right)=-\frac{n^2}{g_{00}} \neq 0;
\end{equation}
being $n$ the refractive index of the medium given by Eqn.\eqref{refra-index} and where the gravitational redshift has already been taken into account,
\begin{equation}
    \omega(x)=\frac{\omega_\infty}{\sqrt{-g_{00}}}.
\end{equation}
In this way we can re-write the Hamiltonian \eqref{hamil-reduce} as follows, 
\begin{equation}\label{eq:hamilredu}
    \hat{\mathcal{H}}=\frac{1}{2}(\hat{g}^{ab}(p_a+\hat{\beta}_a \omega_\infty)(p_b+\hat{\beta}_b \omega_\infty)-\omega_\infty^2),
\end{equation}
where
\begin{equation}
    \hat{g}^{ab}(x)\frac{\partial}{\partial x^a}\frac{\partial}{\partial x^b} =-\frac{g_{00}}{n^2}\bigg( g^{ij}\frac{\partial}{\partial x^i}\frac{\partial}{\partial x^j}+g^{33}(\frac{\partial}{\partial x^3})^2\bigg),
\end{equation}
and with inverse $\hat{g}_{ab}$ (defined as $\hat{g}_{ab}\hat{g}^{bc}=\delta_a^c$) given by,
\begin{equation}\label{eq:opticalmetricf}
    \hat{g}_{ab}=\frac{n^2}{-g_{00}}\bigg( g_{ab}-\frac{g_{0a}g_{0b}}{g_{00}} \bigg).
\end{equation}
To obtain the above expression we have used the identity,
\begin{equation}
    \frac{1}{g^{33}}=g_{33}-\frac{(g_{03})^2}{g_{00}}.
\end{equation}
On the other hand, $\hat{\beta}_a$ are the components of the one-form $\bm{\hat{\beta}}$ given by,
\begin{equation}\label{eq:formbet}
    \bm{\hat\beta}\equiv\hat{\beta}_a(x) dx^a = \frac{g_{03}}{g_{00}}d\varphi.
\end{equation}
It is important to note that $\hat{g}_{ab}$ is a 3-dimensional Riemannian metric and  as follows from the Hamilton equations associated to \eqref{eq:hamilredu} the motion of light rays are determined by \cite{Perlick-book},
\begin{eqnarray}
\hat{g}_{ab}\dot{x}^a\dot{x}^b&=&1, \\\label{geod-hat}
    \ddot{x}^a+\hat{\Gamma}^a_{bc}\dot{x}^b \dot{x}^c &=& \hat{g}^{ad}\left( \partial_e\hat{\beta}_d-\partial_d \hat{\beta}_e \right) \dot{x}^e
\end{eqnarray}
where $\hat{\Gamma}^a_{bc}$ are the Christoffel symbols of the metric $\hat{g}_{ab}$. From \eqref{geod-hat} we can see that the if $\bm{\hat\beta}\neq 0$ light rays of the reduced optical structure in general do not follow geodesics with respect to the metric $\hat{g}_{ab}$. For static spacetimes, they are indeed geodesics of $\hat{g}_{ab}$, and it is known as Jacobi metric. 
    
Finally, the optical path length takes the form,
\begin{equation}\label{functional-I}
    \mathcal{I}=\pm \int_{\tilde{s}_1}^{\tilde{s}_2}\mathcal{F}(x,\dot{x})(\tilde{s}) d\tilde{s},
\end{equation}
where
\begin{equation}\label{Finsler-general}
    \mathcal{F}(x,\dot{x})=\sqrt{\hat{g}_{ab}(x)\dot{x}^a\dot{x}^b}-\hat{\beta}_a(x) \dot{x}^a,
\end{equation}
is a Finsler-Randers type metric. According to the Fermat principle light rays between two points of the reduced ray-optical structure are the extremes of the functional \eqref{functional-I} and in particular they follow geodesics with respect to the Finsler-Randers metric $\mathcal{F}$ (see \cite{bao-Finsler} for a detailed treatment on Finsler geometry).

We want to point out that an explicit expression for the Finsler-Randers metric \eqref{Finsler-general} was obtained for the particular case of photons propagating in a cold non-magnetized plasma over a Kerr spacetime by Perlick in \cite{Perlick-book}. Expressions \eqref{eq:opticalmetricf} and \eqref{eq:formbet} are their generalization to any stationary spacetime, and in fact even when to our knowledge they were not explicitly presented before in the literature,  they are already implicit in \cite{Perlick-book}. On the other hand, it was recently reported in \cite{Gibbons-Jac-Maup-3} an expression of the Finsler-Randers metric for massive particles propagating in a Kerr spacetime. Expressions \eqref{eq:opticalmetricf} and \eqref{eq:formbet} also contain this particular case through the known correspondence between the motion of test massive particles in a given stationary axisymmetric spacetime and photons moving in a homogeneous plasma environment in the same background. We will use this correspondence later to discuss the motion of test massive particles.

\section{Deflection angle in stationary spacetimes using the Gauss-Bonnet theorem}\label{sec-3}

As it is well known, the Gauss-Bonnet theorem connects the intrinsic geometry of a two-dimensional surface with its topology. 

Precisely, it can be enunciated as follows. Let $D\subset S$ be a regular domain of an oriented two-dimensional surface $S$ with a Riemannian metric $\tilde{g}_{ij}$, the
boundary of which is formed by a closed, simple, piece-wise, regular, and positive oriented curve $\partial D: \mathcal{R}\supset I\to D$. Then,
\begin{equation}
    \int\int_D \mathcal{K}dS+\int_{\partial D} k_g dl +\sum_i \epsilon_i = 2\pi\chi(D), \ \ \ \sigma\in I,
\end{equation}
where $\chi(D)$ and $\mathcal{K}$ are the Euler characteristic and Gaussian
curvature of $D$, respectively; $k_g$ is the geodesic curvature of $\partial D$; and $\epsilon_i$ is the exterior angle defined in the ith vertex, in the positive sense.

\begin{figure}[h]
\centering
\includegraphics[clip,width=75mm]{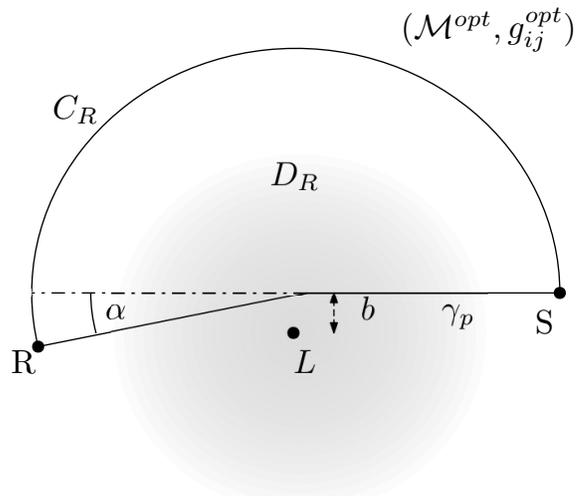}
\caption{The points $S$ and $R$ represent the source and receiver position while $\gamma_p$ indicates a light ray emitted by the source and that reaches the observer. $b$ is identified with the impact parameter and $C_{\text{R}}$ is a semi-circle of radius R connecting $S$ and $R$.}
\label{plot-gb}
\end{figure}

In order to apply the Gauss-Bonnet theorem to calculate the deflection angle of light rays propagating in the equatorial plane in stationary spacetimes we will follow the procedure described first by Gibbons and Werner in \cite{Gibbons-gauss-bonnet} for the pure gravity case and extended  by two of us  to the plasma case \cite{Crisnejo-gauss-bonnet-1}. That is, we will apply the Gauss-Bonnet theorem to a domain $D_R$ as in the Fig. \ref{plot-gb} but at difference with the aforementioned works, we will consider in this case that light rays do not necessarily follow geodesics with respect to the optical metric $g^{\text{opt}}_{ij}$\,\footnote{Later, we will identify this optical metric with the metric $\hat{g}_{ij}$.}. In all the cases we are going to consider it is assumedthat light rays are propagating in the equatorial plane of an stationary axisymmetric spacetime.  This implies in particular that the geodesic curvature of the light rays projected to the optical manifold is not zero. We obtain the following expression for the bending angle $\alpha$,
\begin{equation}
    \int_0^{\pi+\alpha}\bigg[\kappa_g \frac{d\sigma}{d\varphi}\bigg]\bigg|_{C_R}d\varphi=\pi -\bigg(\int\int_{D_R}\mathcal{K}dS + \int_{\gamma_p} k_g dl \bigg),
\end{equation}
where the limit $R\to\infty$ in both sides of the equation is understood.
As was mentioned in the past, for asymptotically flat spacetimes one has that
$[k_g\frac{d\sigma}{d\varphi}]_{C_R}\to 1$ as the radius of $C_R$ goes to infinity which allows to write the deflection angle in an even more simple way,
\begin{equation}\label{alpha-general}
    \alpha=-\int\int_{D_R}\mathcal{K}dS - \int_{\gamma_p} k_g dl.
\end{equation}
In this work we will restrict our attention to asymptotically flat spacetimes and by then the expression \eqref{alpha-general} will be enough to calculate the deflection angle.

We mention that the expression \eqref{alpha-general} is similar to the one obtained by Ono et. al. (see Eqn. (30) in \cite{Ono:2017pie}) where the difference is in the domain of integration because they are computing finite distance corrections to the deflection angle and in the present paper we are not interested in that case.

It has been reported in the literature two different ways to implement the equation \eqref{alpha-general} to calculate the deflection angle in a stationary and axysimmetric spacetime for the vacuum case.  Here, we briefly describe both of them. One method (developed in~\cite{Ono:2017pie}) consists in using the Riemmanian metric $\hat{g}_{ab}$ restricted to the equatorial plane to compute the Gaussian and geodesic curvatures in \eqref{alpha-general} to calculate the bending angle. In that case due to light rays do not follow geodesics with respect to this metric, the geodesic curvature associated with $\gamma_p$ will not be zero. In particular, for light rays moving in the equatorial plane characterized by $\theta=\pi/2$, its geodesic curvature can be calculated as follows,
\begin{equation}\label{eq:kgasada}
    k_g =-\frac{1}{\sqrt{\hat{g} \hat{g}^{\theta\theta} }} \partial_r\hat{\beta}_\varphi,
\end{equation}
where $\hat{g}$ is the determinant of $\hat{g}_{ab}$.

The other method which was reported by Werner in \cite{Werner:2012rc} for stationary spacetimes in the pure gravity case consists in building an osculating Riemannian metric from the Finsler-Randers one by using the Nazim's method \cite{nazim}. The advantage of this method is that the light rays follow geodesics with respect to the resulting Riemannian metric and by them the second term in \eqref{alpha-general} vanishes. The disadvantage is that the computations are more cumbersome even at linear order.   

In the following two sections we will compute the deflection angle of light rays propagating in a plasma environment from the equation \eqref{alpha-general} in the weak field approximation.

\section{Deflection angle in weak field approximation using Ono {\it et.al.} approach}\label{sec-4}
We are interested in the computation of the deflection angle in the weak field regime for an stationary and axisymmetric spacetime surrounded by a cold-plasma medium using the method proposed in \cite{Ono:2017pie}. In particular, in this Section we will obtain a general expression for the deflection angle taking into account first order corrections. That is, we will consider an stationary metric written as
\begin{equation}\label{eq:metricoriginal}
\begin{split}
    ds^2&=-(1+\epsilon_1 h_{tt}(r,\theta))dt^2+(1+\epsilon_2 h_{rr}(r,\theta))dr^2\\
    &+2\epsilon_3 h_{t\varphi}dtd\varphi +r^2(d\theta^2+\sin^2\theta d\varphi^2),
    \end{split}
\end{equation}
and we will keep only those terms which are linear in the bookkeeping parameters $\epsilon_1$, $\epsilon_2$ and $\epsilon_3$.
In general is assumed that $h_{t\varphi}$ is proportional to a spin parameter $a$.
We will restrict all the discussion to the equatorial plane $\theta=\pi/2.$
At the considered order the Finsler-Randers metric is determined by
\begin{eqnarray}
    \hat{g}_{ab} dx^a dx^b&=&n^2[(1-\epsilon_1 h_{tt}+\epsilon_2 h_{rr})dr^2\nonumber\\
    &&+r^2(1-\epsilon_1h_{tt})d\varphi^2],\nonumber\\
    \bm{\hat\beta}&=&\epsilon_3 h_{t\varphi}d\varphi.
\end{eqnarray}

For a cold non magnetized plasma with a charge density profile of the form $N(r)=N_0+N_1(r)$, with $N_0=\text{constant}$, it follows that the refraction index can be written as
\begin{equation}
    n^2=1-\frac{K_e N(r)}{\omega_\infty^2}(1+\epsilon_1 h_{tt}),
\end{equation}
where we have used that $\omega^2_e=K_eN(r)=\omega^2_{e0}+K_eN_1(r)$. 
The deflection angle is determined by:
\begin{equation}\label{eq:asa}
\alpha=\underbrace{-\int\int_{D_r}\mathcal{K}dS}_{\alpha_{\mathcal{K}}}\underbrace{-\int^S_R \kappa_g dl}_{\alpha_{k_g}}.
\end{equation}
Let us start with the computation of the Gaussian curvature at the considered order. It is given by:
\begin{equation}\label{eq:kds1o}
   \mathcal{K}dS=(\epsilon_2\tilde{\mathcal{K}}_{\epsilon_2}+\epsilon_1\tilde{\mathcal{K}}_{\epsilon_1}+\tilde{\mathcal{K}}_{\text{plasma}})drd\varphi; 
\end{equation}
with
\begin{eqnarray}
\tilde{\mathcal{K}}_{\epsilon_2}&=&\frac{d}{dr} \bigg(\frac{h_{rr}}{2}\bigg),\\
\tilde{\mathcal{K}}_{\epsilon_1}&=&\frac{d}{dr}\bigg(\frac{\omega_\infty^2 r h_{tt}^\prime}{2(\omega_\infty^2-\omega_{e0}^2)}\bigg),\\
\tilde{\mathcal{K}}_{\text{plasma}}&=&\frac{d}{dr}\bigg(\frac{K_e r N_1^\prime}{2(\omega_\infty^2-\omega_{e0}^2)}\bigg).
\end{eqnarray}
Let us assume that $\tilde{\mathcal{K}}_{\epsilon_2},$ $\tilde{\mathcal{K}}_{\epsilon_1}$ and $\tilde{\mathcal{K}}_{\text{plasma}}$ are different from zero, i.e., neither $h_{rr}$, $r h_{tt}^\prime$ or ${r N_1^\prime}$ take constant values. In the following and the rest of the paper, we will only consider prograde orbits, that is orbits whose orbital angular momentum are in the same direction as the spin of the considered metric.

 As we are interested in the first order contribution to the deflection angle it is enough to integrate along a domain $D_r$ where the curve $\gamma_p$ is approximated by the path followed by light rays in the flat background, that is, straight lines. This is called the Born approximation. Therefore the equation for $\gamma_p$ in the integration domain $D_r$ can be replaced by the flat geodesic straight line given by $r_\varphi=\f{b}{\sin\varphi}$, while the polar angular coordinate $\varphi$ runs from $0$ to $\pi$. 

In that situation, the integration of \eqref{eq:kds1o} gives
\begin{equation}
    \alpha_{\mathcal{K}}=\frac{1}{2}\int^\pi_0 \bigg(\epsilon_2 h_{rr}+\epsilon_1 \frac{\omega_\infty^2 r h_{tt}^\prime}{\omega_\infty^2-\omega_{e0}^2}+\frac{r K_e  N_1^\prime}{\omega_\infty^2-\omega_{e0}^2}\bigg)\bigg|_{r_\varphi} d\varphi.
\end{equation}
  As we will see in Sec. \ref{third-order}, for higher order corrections this approximation should be improved.

Let us see $\alpha_{k_g}$,
\begin{equation}
    \alpha_{k_g}= \int_0^\pi \bigg(k_g \frac{dl}{d\varphi}\bigg)\bigg|_{r_\varphi} d\varphi.
\end{equation}
The geodesic curvature $k_g$ obtained from \eqref{eq:kgasada} is given by,
\begin{equation}
    k_g=-\epsilon_3\frac{h_{t\varphi}^\prime}{r}\frac{\omega_\infty^2}{\omega_\infty^2-\omega_{e0}^2}+\mathcal{O}(\epsilon_3\epsilon_1,\epsilon^2_3).
\end{equation}
On the other hand for prograde orbits,
\begin{equation}
    \frac{dl}{d\varphi}=\frac{\sqrt{\omega_\infty^2-\omega_{e0}^2}}{\omega_\infty b}r^2\bigg|_{r_\varphi}.
\end{equation}
Then,
\begin{equation}
    \alpha_{k_g}=- \int_0^\pi \frac{\epsilon_3}{b} \bigg(\frac{r \omega_\infty h_{t\varphi}^\prime}{\sqrt{\omega_\infty^2-\omega_{e0}^2}}\bigg)\bigg|_{r\varphi}d\varphi.
\end{equation}
Finally the following expression for the deflection angle results,
\begin{equation}\label{eq:signoi}
\begin{aligned}
    \alpha=\frac{1}{2}\int^\pi_0 \bigg(&\epsilon_2 h_{rr}+\epsilon_1 \frac{\omega_\infty^2 r h_{tt}^\prime}{\omega_\infty^2-\omega_{e0}^2}+\frac{r K_e  N_1^\prime}{\omega_\infty^2-\omega_{e0}^2} \\
    -&\epsilon_3\frac{2r \omega_\infty h_{t\varphi}^\prime}{b\sqrt{\omega_\infty^2-\omega_{e0}^2}}  \bigg)\bigg|_{r_\varphi} d\varphi.
\end{aligned}    
\end{equation}

For static spacetimes ($\epsilon_3=0$) this expression can be compared to the one obtained by Bisnovatyi-Kogan and Tsupko in \cite{BisnovatyiKogan:2010ar} (see Eqn. (30) on that reference). The expression \eqref{eq:signoi} is given in radial coordinates instead of the Cartesian ones used in \cite{BisnovatyiKogan:2010ar} but after a change of coordinates and integration by part it can be shown that they are completely equivalent.

The case of retrograde orbits can be analyzed in a similar way under a correspondent domain $D'_r$, resulting in a similar expression to \eqref{eq:signoi} but with an opposite sign in the last term. Alternatively, it can be understood as a result that the motion of a retrograde photon (with orbital angular momentum $p_\varphi<0$) in the original coordinate expression for the metric (with positive spin $a$ and starting in the asymptotic region in $\varphi_S=0$ and with the receiver position at $\varphi_R=-\pi$) is equivalent to the motion of a photon with positive angular momentum $p_\varphi>0$ in a metric of the same coordinate form as \eqref{eq:metricoriginal} but obtained by replacing $a$ by $-a$ and studying orbits starting in the asymptotic region in $\varphi_S=0$ and ending in $\varphi_R=\pi$.

{It is worthwhile to note that unlike the exact expression for the deflection angle  in terms of the closest approach distance in the Kerr spacetime for non-homogeneous plasma derived by Perlick in \cite{Perlick-book}, in the practice it is usually more convenient to work with an approximate expression in terms of the impact parameter instead of the closest approach distance because it is a parameter at infinity. Therefore, for applications in the weak-field approximation the expression \eqref{eq:signoi} is more convenient to the one obtained by Perlick in \cite{Perlick-book}. Of course, the expression for the deflection angle described in \cite{Perlick-book} is not limited to the weak field regime, and it must be full considered in the case of strong deflection.  In Sec. \ref{sec-6} we will derive an expression for the deflection angle with the same characteristics of \eqref{eq:signoi} in the Kerr spacetime and for different electronic density profiles taking into account higher order corrections.}

\subsection{Correspondence between homogeneous plasma and massive particles}\label{corresp}
As is well known there exist a correspondence between the motion of a photon in a cold non magnetized plasma and the motion of a test massive particle in the same background \cite{Kulsrud-1992}.
This correspondence is obtained by identifying the electronic frequency of the plasma  $\hbar\omega_e$ with the mass $m$ of the test massive particle and the total energy $E=\hbar \omega_\infty$ of a photon with the total energy $E_\infty=m/(1-v^2)^{1/2}$.
 With this identification at hand, and by setting the bookkeeping parameters as $\epsilon_2=\epsilon_1=\epsilon_3=1$, we can express the deflection angle for test massive particles as follows,
\begin{equation}\label{alpha-pm}
    \alpha_{\text{mp}}=\frac{1}{2}\int_0^\pi \bigg( h_{rr}+ \frac{r h_{tt}^\prime}{v^2}- 2s\frac{r h_{t\varphi}^\prime}{bv}\bigg)\bigg|_{r_\varphi}d\varphi,
\end{equation}
with $s=+1$ for prograde orbits and $s=-1$ for retrograde ones.
As applications of this expression in the next subsection we will consider the propagation of a test massive particle in a Kerr-Newmann background and in a rotating Teo wormhole.

\subsubsection{Kerr-Newman spacetime}
Let us consider a Kerr-Newman spacetime given by the following line element at first order in $a,M$ and $Q^2$,
\begin{equation}
\begin{aligned}
    ds^2=& -(1-\frac{2M}{r}+\frac{Q^2}{r^2})dt^2-\frac{4aM\sin^2\theta}{r} dt d\varphi \\
    &+ (1+\frac{2m}{r}-\frac{Q^2}{r^2})dr^2 + r^2(d\theta^2+\sin^2\theta d\varphi^2) \\
    &+ \mathcal{O}(M^2,a^2,aQ^2).
\end{aligned}    
\end{equation}
By restricting our attention to the equatorial plane $\theta=\pi/2$ we have,
\begin{eqnarray}
h_{tt}(r)&=&-\frac{2M}{r}+\frac{Q^2}{r^2}, \\
h_{rr}(r)&=&\frac{2M}{r}-\frac{Q^2}{r^2}, \\
h_{t\varphi}(r)&=&-\frac{2aM}{r}.
\end{eqnarray}

Using the expression \eqref{alpha-pm} we obtain the deflection angle for prograde/retrograde orbits of massive particles,
\begin{equation}\label{eq:eRN}
    \alpha_{\text{mp}} = \frac{2M}{b}\bigg(1+\frac{1}{v^2}\bigg)-\frac{\pi Q^2}{4b^2}\bigg(1+\frac{2}{v^2}\bigg)- \frac{4saM}{b^2 v},
\end{equation}
which coincides with the one obtained in \cite{Jusufi:2019rcw} for uncharged particles. 

\subsubsection{A rotating Teo wormhole}
As a second application, let us consider a rotating Teo wormhole described by the line element,

\begin{equation}\label{Teo-1}
ds^2=-N^2 dt^2+\frac{dr^2}{1-\frac{b_0}{r}}+r^2 K^2\left[d\theta^2+\sin^2\theta(d\varphi-\omega dt)^2  \right] 
\end{equation}
where 
\begin{equation}
N=K=1+\frac{(4 a \cos\theta)^2}{r},
\end{equation}
and
\begin{equation}
\tilde{\omega}=\frac{2 a}{r^3}.
\end{equation}
In order to apply the expression \eqref{alpha-pm} to calculate the deflection angle of massive particles at the equatorial plane in this spacetime, it is enough to consider the line element \eqref{Teo-1} at first order in the parameters $b_0$ and $a$:
\begin{equation}
    ds^2=-dt^2+(1+\frac{b_0}{r})dr^2+r^2 d\varphi^2 -\frac{4a}{r}dt d\varphi + \mathcal{O}(b_0^2,a^2).
\end{equation}

In this case we can identify,
\begin{equation}
h_{tt}=0,\ \ \ h_{rr}(r)=\frac{b_0}{r}, \ \ \ h_{t\varphi}(r)=-\frac{2a}{r}.
\end{equation}
 By using the expression \eqref{alpha-pm} for the deflection angle of prograde/retrograde orbits we finally obtain,
\begin{equation}\label{eq:teoangle}
    \alpha_{\text{mp}} = \frac{b_0}{b}- \frac{ 4sa}{b^2 v} +\mathcal{O}(\frac{b_0^2}{b^2},\frac{a^2}{b^2}),
\end{equation}
which coincides with the one obtained in \cite{Jusufi:2018kry}.

\section{Deflection angle in weak field approximation using Werner's approach: A case of massive particles}\label{sec-5}
We show here that an alternative method proposed by Werner to compute the deflection angle of light rays in stationary spacetimes, can also be used to describe the motion of test massive particles. Due to this method needs the non trivial construction of an osculating Riemannian metric even in the situation of a linearized and slowly rotating source is a cumbersome task to derive an equation like \eqref{alpha-pm}. For this reason, using this approach we will show its effectiveness by recovering the results previously presented for the Kerr spacetime. 
\subsection{Kerr spacetime}
Let us consider the Kerr metric in the equatorial plane $\theta=\pi/2$ given by
\begin{eqnarray}\notag
ds^2&=&-\left(1-\frac{2M}{r}\right)dt^2+\frac{r^2}{\Delta}dr^2- \frac{4 aM }{r}dt d\varphi\\
&+& \left(r^2+a^2+\frac{2Ma^2}{r}\right) d\varphi^2 \label{metric}
\end{eqnarray}
where 
\begin{equation}
\Delta=r^2-2 M r+a^2.
\end{equation}

In what follows we will use  the metric (\ref{metric}) to find out the  deflection angle of of massive particles. Towards this purpose let us find the corresponding Finsler-Randers metric for the Kerr metric. It is well known that a Finsler-Randers metric $\mathcal{F}$ with manifold $\mathcal{M}$ and $x\in \mathcal{M},\ X\in T_x M$ is characterised by the Hessian given as \cite{Werner:2012rc}
\begin{equation} 
g_{ij}(x,X)=\frac{1}{2}\frac{\partial^{2}F^{2}(x,X)}{\partial X^{i}\partial X^{j}}.\label{10-3}
\end{equation}

Using the correspondence between the motion of photons in a homogeneous plasma and the motion of test massive particles as discussed in the subsection \eqref{corresp} we can study the propagation of test massive particles over the Kerr spacetime.

In particular the particle of mass $m$ is assumed leaving the asymptotic
region with a speed $v$ as measured by an asymptotic observer and therefore with an energy
\begin{equation}\label{eq:energyinf}
E_\infty=\frac{m}{\sqrt{1-v^2}}.
\end{equation}
In the same way let us assume that the particle has an
angular momentum
\begin{equation}\label{eq:angumom}
J=\frac{m v b}{\sqrt{1-v^2}}.
\end{equation}
As before, our lensing setup is adapted for prograde orbits. Then, we can consider the massive particle as propagating through an effective medium with refraction index given by,
\begin{equation}\label{eq:refindx}
n^2(r)=1-\frac{m^2}{E^2}(1-\frac{2M}{r})=1-(1-v^2)\left(1-\frac{2M}{r}\right).
\end{equation}

The associated Finsler-Randers type metric reads,
\begin{eqnarray}\notag
&\mathcal{F}&\left(r,\varphi,\frac{\mathrm{d}r}{\mathrm{d}t},\frac{\mathrm{d}\varphi}{\mathrm{d}t}\right)=\left[1-(1-v^2)\left(1-\frac{2M}{r}\right)\right]^{1/2}\\\notag
&\times& \left[\frac{r^4\Delta }{(\Delta-a^2)^2}\left(\frac{\mathrm{d}\varphi}{\mathrm{d}t}\right)^2+\frac{r^4 }{\Delta(\Delta-a^2)}\left(\frac{\mathrm{d}r}{\mathrm{d}t}\right)^2\right]^{1/2}\\
&-&\frac{2Mar}{\Delta-a^2}\frac{\mathrm{d}\varphi}{\mathrm{d}t}.
\end{eqnarray}

As we have said before, according to the Fermat's principle in general relativity test massive particles (due to the correspondence with photons in a homogeneous plasma) follow geodesics curves with respect to the Finsler-Randers type metric $\mathcal{F}$. That is, test massive particles follow geodesics given by the condition, 
\begin{equation}
0=\delta\,\int\limits_{\gamma_\mathcal{F}}\mathcal{F}(x, \dot{x})\mathrm{d}t.
\end{equation}

It is worth noting that $\gamma_\mathcal{F}$ represent a geodesic of the Kerr-Randers optical metric $\mathcal{F}$. Following Werner's method \cite{Werner:2012rc}, one can now construct a Riemannian manifold $(\mathcal{M},\bar{g})$ which osculates the Randers manifold $ (\mathcal{M}, \mathcal{F}) $ using the so-called Naz{\i}m's method \cite{nazim}. In particular, one can choose a vector field $\bar{X}$ tangent to the geodesic $\gamma_{\mathcal{F}}$, with $\bar{X}(\gamma_{\mathcal{F}})=\dot{x}$, and the new metric characterised by the Hessian 
\begin{equation}
\bar{g}_{ij}(x)=g_{ij}(x,\bar{X}(x)).\label{17-3}
\end{equation}

At this point, we recall the striking property according to which  the geodesic $\gamma_{\mathcal{F}}$ of the Finsler-Randers manifold is also a geodesic $\gamma_{\bar{g}}$ of $(\mathcal{M},\bar{g})$, in other words $\gamma_{\mathcal{F}}=\gamma_{\bar{g}}$ ( for details see \cite{Werner:2012rc}). In what follows we are going to use the osculating Riemannian manifold $(\mathcal{M},\bar{g})$  to compute the deflection angle of a test massive particle. At leading order, it suffices to use the Born approximation for the path of light ray  ($r(\varphi)=b/\sin\varphi$).  %where $b$ is the impact parameter and can be approximated as the minimal distance to the black hole. 
Near the light ray, one can choose the vector field as follows
\begin{eqnarray}\label{vec}
\bar{X}^{r}&=&-\cos\varphi+\mathcal{O}(a,M), \\
\bar{X}^{\varphi}&=&\frac{\sin^{2}\varphi}{b}+\mathcal{O}(a,M).
\end{eqnarray}

As shown in the next subsection for asymptotically flat spacetimes the deflection angle will be given by,
\begin{equation}\label{alpha-werner}
    \alpha_{\text{mp}}=-\int\int_{D_R}\mathcal{K}dS,
\end{equation}
where $\mathcal{K}$ and $dS$ are the Gaussian and surface element associated with the osculating optical metric $\bar{g}$.

\subsubsection{Gaussian optical curvature and deflection angle}

Let us now proceed to use of the osculating Riemannian manifold $(D_{R},\bar{g})$ in a domain $(\mathcal{M},\bar{g})$ with the boundary $\partial D_{R}=\gamma_{\bar{g}}\cup C_ {R}$. In that case, one can apply the GBT stated as
\begin{equation} \label{gb2}
\iint\limits_{D_{R}}\mathcal{K}\,\mathrm{d}S+\oint\limits_{\partial D_{R}}\kappa\,\mathrm{d}t=2\pi\chi(D_{R})-(\theta_{R}+\theta_{S})=\pi,
\end{equation} 
with $\mathcal{K}$ being the optical Gaussian curvature and $\kappa=|\nabla_{\dot{\gamma}}\dot{\gamma}|$ the geodesic curvature. Furthermore we have used  the fact that the sum of exterior jump angles at $S$ and $R$ satisfies $\theta_{R}+\theta_{S}\to \pi$ and the Euler characteristic $\chi(D_{R})=1$. 
In this way, letting $R\to \infty$ the GBT reduces to the following form
\begin{eqnarray}
\iint\limits_{D_{R}}\mathcal{K}\,\mathrm{d}S & + & \oint\limits_{C_{R}}\kappa\,\mathrm{d}t\overset{{R\to \infty}}{=}\iint\limits_{
D_{\infty}}\mathcal{K}\,\mathrm{d}S + \int\limits_{0}^{\pi+ \hat{\alpha}}\mathrm{d}\varphi,
\end{eqnarray}
integrated over the optical domain $D_\infty$ outside the light ray $\gamma_{\bar{g}}$. 
In addition we have used the fact that since $\gamma_{\bar{g}}$ is geodesic one must have $\kappa(\gamma_{\bar{g}})=0$. Thus in the limit  $R\to \infty$ it is not difficult to show that the geodesic curvature yields 
\begin{equation}
\kappa(C_{R})\to \frac{1}{v R}. 
\end{equation}
While for a constant radial coordinate the optical metric yields
\begin{eqnarray}\notag
\mathrm{d}t&=&\Big[\sqrt{\left[1-(1-v^2)\left(1-\frac{2M}{R}\right)\right]\frac{R^4\Delta }{(\Delta-a^2)^2}}\\
&-&\frac{2aM  R}{\Delta-a^2}\Big]\mathrm{d}\varphi\to  v R \mathrm{d}\varphi.
\end{eqnarray}
Combining these expressions we find 
\begin{eqnarray}
\lim_{R \to \infty} \left[\kappa(C_{R})\frac{\mathrm{d}t}{\mathrm{d} \varphi}\right]_{C_{R}}	 \to 1.
\end{eqnarray}
In other words, the deflection angle (66) is immediately obtained.

Finally we can compute the corresponding metric components $\bar{g}$ of the osculating Riemannian geometry as well as the Gaussian optical curvature $\mathcal{K}$. Applying the Hessian (\ref{10-3}), along with the vector field  Eqn. (\ref{vec}), we find the following expressions
\begin{widetext}
\begin{eqnarray}
\bar{g}_{rr}&=&v^2+\frac{2(1+v^2)M}{r}-\frac{2 Ma r v \sin^6\varphi}{\left(r^2  \sin^4 \varphi+\cos^2 \varphi \,b^2 \right)^{3/2}}+\mathcal{O}(\frac{M^2}{b^2},\frac{a^2}{b^2}),\\
\bar{g}_{\varphi \varphi}&=&r^2 v^2+2Mr-\frac{2 M a v \sin^2\varphi r (2 \sin^4\varphi r^2+3 \cos^2 \varphi b^2)}{\left(r^2  \sin^4 \varphi+\cos^2 \varphi \,b^2 \right)^{3/2}}+\mathcal{O}(\frac{M^2}{b^2},\frac{a^2}{b^2})\\ 
\bar{g}_{r\varphi}&=&\frac{2 a v M  \cos^3\varphi}{r \left(\frac{r^2 \sin^4 \varphi+\cos^2 \varphi \,b^2 }{b^2}\right)^{3/2}}+\mathcal{O}(\frac{M^2}{b^2},\frac{a^2}{b^2}),
\end{eqnarray}
and the determinant given by
\begin{equation}
\det \bar{g}= r^2v^2 +2 v^2 M r (v^2+2)-\frac{6 Ma v^3 r \sin^2\varphi}{\sqrt{r^2  \sin^4 \varphi+\cos^2 \varphi \,b^2}}+\mathcal{O}(a^2,M^2).
\end{equation}

Indeed one can easily observe that in the limit $v=1$, these equations reduces to light deflection case \cite{Werner:2012rc}. 
The  Gaussian optical curvature is defined by the following relation
\begin{eqnarray}\label{25}
\mathcal{K}=\frac{\bar{R}_{r\varphi r\varphi}}{\det \bar{g}}=\frac{1}{\sqrt{\det \bar{g}}}\left[\frac{\partial}{\partial \varphi}\left(\frac{\sqrt{\det \bar{g}}}{\bar{g}_{rr}}\,\bar{\Gamma}^{\varphi}_{rr}\right)-\frac{\partial}{\partial r}\left(\frac{\sqrt{\det \bar{g}}}{\bar{g}_{rr}}\,\bar{\Gamma}^{\varphi}_{r\varphi}\right)\right].
\end{eqnarray}

Making use of the above expression we find the following result in leading order terms
\begin{eqnarray}\label{k1}
\mathcal{K}&=& - \frac{M\,(v^2+1)}{v^4 r^3}+\frac{M \,a }{r^2 v^3 }f(r,\varphi)+\mathcal{O}(\frac{M^2}{b^4},\frac{a^2}{b^4})
\end{eqnarray}
where
\begin{eqnarray}\label{k2}\notag
f(r,\varphi)&=& \frac{1}{\left(r^2  \sin^4 \varphi+\cos^2 \varphi \,b^2 \right)^{7/2}}\Big(30 \cos^4\varphi \sin^8\varphi b^2 r^3-6\sin^{14}\varphi r^5+12 \cos^2\varphi \sin^{10}\varphi b^2 r^3\\\notag
&-& 48 \cos^4\varphi \sin^7 \varphi b^3 r^2-24 \cos^2 \varphi \sin^9 \varphi b^3 r^2-30\cos^6\varphi \sin^4\varphi b^4 r-27 \cos^4\varphi \sin^6 \varphi b^4 r \\
&-& 12 \cos^2\varphi \sin^8\varphi b^4 r+12 \cos^6\varphi \sin^3\varphi b^5+6 \cos^4\varphi \sin^5\varphi b^5 \Big).
\end{eqnarray}
\end{widetext}

Using the equations \eqref{k1} and \eqref{k2} and following \eqref{alpha-werner} the deflection angle can be expressed as follows,
\begin{equation}\label{eq:alphaftheta}
\alpha_{\text{mp}} = - \int\limits_{0}^{\pi}\int\limits_{\frac{b}{\sin \varphi}}^{\infty}\left(- \frac{M(v^2+1)}{v^2 r^2}+\frac{M a }{r v }f(r,\varphi)\right)\mathrm{d}r\,\mathrm{d}\varphi.
\end{equation}
Finally by evaluating the above integral we find
\begin{equation} \label{eq:alp2}
\alpha_{\text{mp}} = \frac{2M}{b}\bigg(1+\frac{1}{v^2}\bigg) - \frac{ 4 M a}{ b^2 v},
\end{equation}
which coincides with \eqref{eq:eRN} by setting $Q=0$.
In the limit $v=1$, we do recover the deflection angle of light by a Kerr black hole \cite{Edery2006,doi:10.1063/1.1705193,Aazami:2011tw}. %
In this way we show that it is also possible to calculate the deflection angle of massive particles using the Werner's method, or equivalently to study the motion of photons in a homogeneous plasma. In order to study the propagation of photons in a non-homogeneous plasma we prefer to apply the method discussed in section \ref{sec-4} because it is more direct. We are going to study this case in Section \ref{third-order}.

 For completeness, and in order to show how Werner's method is a bit more cumbersome to obtain the deflection angle, in Appendix \eqref{app} we apply it again to calculate the deflection angle in the rotating Teo spacetime.

\section{Higher order corrections to the deflection angle}\label{sec-6}

\subsection{Deflection angle of massive particles up to third order in a Kerr background}\label{third-order}

  As we have shown, at linear order the expression \eqref{alpha-pm} is a general formula that follows from the use of the Gauss-Bonnet method, and therefore at this order it is not necessary to repeat the computations of the Gaussian and geodesic curvatures for each one of the linearized metrics under study. 
 
 However, in order to compute higher order corrections a general formula is not so convenient. Moreover, at higher order, the Born approximation will not be enough, and the integration domain will explicitly depend on the details of the orbit. To our best knowledge, all the existing computations for the deflection angle in stationary spacetimes based in the Gauss-Bonnet theorem have been limited to find linear order corrections due to the intrinsic spin of these metrics (See \cite{Ono:2018ybw,Ono:2018jrv,Jusufi:2017mav} to cite some examples).
 
 Now we will procedure to fill this gap by computing the deflection angle for test relativistic massive particles in a Kerr background including third order corrections, that is containing terms of higher order as $\frac{M^3}{b^3},\frac{M^2a}{b^3}$ and $\frac{M a^2}{b^3}$. In particular, we will see that for the case of massless particles our general expression reduces to a known formula obtained in the past using other techniques \cite{Iyer:2009hq,Aazami:2011tw}.

As mentioned above, at higher order we need to go beyond the Born approximation. In particular, in order to compute third order corrections, we need to know the orbit of the massive/massless particle at second order. 

Before to do that let us write the general expression for the orbit in the equatorial plane of a general stationary spacetime given by the metric:
\begin{equation}\label{eq:equati}
ds^2|_{\theta=\frac{\pi}{2}}=-Adt^2-2Hdtd\varphi+Bdr^2+Dd\varphi^2.
\end{equation}
From the Hamiltonian $\mathcal{H}=\frac{1}{2}(g^{\alpha\beta}p_\alpha p_\beta+m^2)$, it follows that the orbit equation for the prograde orbit of a  particle which is assumed leaving the asymptotic
region with a speed $v$ as measured by an asymptotic observer and therefore with an energy and angular momentum given by Eqn.\eqref{eq:energyinf} and \eqref{eq:angumom} is given by
\begin{equation}\label{eq:F}
\bigg(\frac{du_\gamma}{d\varphi}\bigg)^2=\frac{u^4_\gamma\Delta}{(H+Abv)^2B}[-\Delta(1-v^2)+D-2Hbv-Ab^2v^2)];
\end{equation}
with $\Delta=AD+H^2$ and where $u_\gamma=1/r_\gamma$. For massless particles ($v=1$) it reduces to the expression given in \cite{Ono:2017pie}.

Let us concentrate on the Kerr spacetime, for other stationary metrics the procedure will be similar. In that case the metric components at the equatorial plane are given by:
\begin{eqnarray}
A&=&1-\frac{2M}{r},\label{ew:a1}\\
H&=&\frac{2aM}{r},\\
B&=&\frac{r^2}{\Delta},\\
D&=&r^2+a^2+\frac{2a^2M}{r}\label{ew:a4};
\end{eqnarray}
with $\Delta=r^2+a^2-2 M r$.

In this case, the orbit equation reduces to:
\begin{equation}\label{eq:itera}
\begin{split}
    \bigg(\frac{du_\gamma}{d\varphi}\bigg)^2=&\frac{1+a^2u^2_\gamma-2 M u_\gamma}{[2M u_\gamma(a-bv)+bv]^2}\bigg[2M(a-bv)^2u^3_\gamma\bigg.\\
    \bigg.&+v^2(a^2-b^2)u^2_\gamma+2M(1-v^2)u_\gamma+v^2\bigg].
    \end{split}
\end{equation}
As we are working in the weak regime, let us  consider the bookkeeping parameters
\begin{equation}\label{eq:bookpar}
    \gamma=\frac{M}{b}\ll 1; \,\,\,\,\,\,\delta=\frac{a}{b}\ll 1.
\end{equation} 
As we are interested in an expression for the deflection angle which be correct up to third order in these parameters, that is an expression containing terms of the form $\gamma^3, \gamma^2\delta$ and $\gamma\delta^2$,
we need to know $u_\gamma$ at second order in these parameters, that is, we require an expression for $u$ of the form:
\begin{equation}
    u_\gamma=u_0+u_1\gamma+u_2\delta+u_3\gamma^2+u_4\delta\gamma+u_5\delta^2+\mathcal{O}(\gamma^3,\gamma\delta^2,\delta\gamma^2,\delta^3).
\end{equation}
By putting this ansatz into Eqn.\eqref{eq:itera}, and by imposing the asymptotic condition that $\lim_{\varphi\to 0}u=0$, results in the following iterative solution for $u_\gamma$:
\begin{eqnarray}
\begin{aligned}
u_0=&\frac{\sin\varphi}{b},\\
u_1=&\frac{(\cos\varphi-1)(v^2\cos\varphi-1)}{bv^2},\\
u_2=&0,\\
u_3=&-\frac{1}{16bv^2}\bigg[12\varphi\cos\varphi(4+v^2)+3v^2\sin(3\varphi)\bigg.\\\bigg.&+(11v^2-16)\sin\varphi-16\sin(2\varphi)(1-v^2)\bigg]\\
u_4=&-\frac{2(1-\cos\varphi)}{bv},\\
u_5=&\frac{\sin^3\varphi}{2b}.
\end{aligned}
\end{eqnarray}
The fact that $u_5$ (proportional to $\f{a^2}{b^2}$) is different from zero even for a flat spacetime ($M=0$) should be expected because the coordinates $r,\varphi$ are spheroidal instead or spherical, and therefore at order $a^2$ in this coordinates the straight geodesic line is not given by $r=b/\sin\varphi$. For the same reason, the fact that $u_2$ is zero (proportional to $\frac{a}{b}$), should also be expected because when $M=0$,  the correspondent flat metric depends on $a^2$, and therefore at linear order in $a$ the coordinates behave as the standard polar coordinates.

From \eqref{eq:opticalmetricf} we obtain the associated optical metric:
\begin{eqnarray}
\hat{g}_{rr}&=&\frac{n^2r^4} {(r^2-2M r)\Delta},\\
\hat{g}_{\varphi\varphi}&=&\frac{n^2r^2\Delta}{(r-2M)^2}.
\end{eqnarray}
with the refractive index given by Eqn.\eqref{eq:refindx}.  

Now we proceed to compute the deflection angle using Eqn.\eqref{eq:asa}. Before to do that, we must to define the integration domain. This domain will be bounded from below by the orbit under analysis determined by $r_\gamma\equiv1/u_\gamma$ and $r=\infty$ from above. For the angular coordinate $\varphi$, we have set the source position at $\varphi_S=0$, and for the position of the observer we must to set $\varphi_R=\pi+\alpha^{(1)}$, with $\alpha^{(1)}$ the first order correction in the mass to the deflection angle, that is, in our case given by the first term of Eqn.\eqref{eq:eRN}. The reason that second order contributions to $\alpha$ are not needed to add to $\varphi_R$ can be found in \cite{Crisnejo-finite}. In fact, in that reference we show that to obtain second order constructions to the deflection angle the approximation $\varphi_R=\pi$ was enough. We will back to this issue later.  

At the consider order $\mathcal{K}dS$ reads;
\begin{equation}
\begin{split}
    \mathcal{K}dS=&\bigg[\underbrace{\bigg(1+\frac{1}{v^2}\bigg)M}_{\mathcal{\hat{K}}_M}\underbrace{+\bigg(1+\frac{6}{v^2}-\frac{4}{v^4}\bigg)M^2u}_{\mathcal{\hat{K}}_{M^2}}\bigg.\\
    &\bigg.\underbrace{+\frac{3}{2}\bigg(1+\frac{15}{v^2}-\frac{20}{v^4}+\frac{8}{v^6}\bigg)M^3u^2}_{\mathcal{\hat{K}}_{M^3}}\bigg.\\
    &\bigg.\underbrace{+3\bigg(1+\frac{1}{v^2}\bigg)a^2 M u^2}_{\mathcal{\hat{K}}_{Ma^2}}\bigg]dud\varphi,
    \end{split}
\end{equation}
%\end{widetext}
and therefore the contribution to the deflection angle in terms of $u_\gamma$ reads
\begin{equation}\label{eq:alphaK}
\begin{split}
    \alpha_{\mathcal{K}}=&\int^{\pi+\frac{2M}{b}(1+\frac{1}{v^2})}_0\int^{u_\gamma}_0 (\mathcal{\hat{K}}_M+\mathcal{\hat{K}}_{M^2}+\mathcal{\hat{K}}_{M^3}+\mathcal{\hat{K}}_{Ma^2})dud\varphi\\
    =&\frac{2M}{b}(1+\frac{1}{v^2})+\frac{3\pi(4+v^2)}{4v^2}\frac{M^2}{b^2}\\
    &+\frac{2(15v^2+5v^6+45v^4-1)}{3v^6}\frac{M^3}{b^3}\\
    &-\frac{2\pi(1+v^2)}{v^3}\frac{aM^2}{b^3}+\frac{2(1+v^2)}{v^2}\frac{a^2M}{b^3}\\
    &+\mathcal{O}(\frac{M^4}{b^4},\frac{M^3a}{b^4},\frac{M^2a^2}{b^4},\frac{Ma^3}{b^4}).
    \end{split}
\end{equation}
Let us remark that in the integration of the angular variable, appear some trigonometric functions that must be valued in $\varphi_R$, and therefore they must be re-expanded in terms of $\gamma$ and $\delta$ around $\pi$. As anticipated above, it can be checked by direct computation that if instead of taking $\varphi_R=\pi+\alpha^{(1)}$ we take $\varphi_R=\pi+\alpha^{(1)}+\alpha^{(2)}$, (with $\alpha^{(2)}$ been formed by terms of order $\delta^2=M^2/b^2$ and $\delta\gamma=Ma/b^2$)) the contribution to $\alpha_{\mathcal{K}}$ of the resulting new terms are of higher than third order. It is easily seen from the fact that the only terms that could potentially contribute at third order to the deflection angle (by introducing second order corrections in $\varphi_R$)  arise from the integration in the angular variable of \begin{equation} \label{eq:intk}
\int^{u_\gamma}_0\tilde{\mathcal{K}}_Mdu=(1+\f{1}{v^2})\f{M}{b}\sin\varphi+\mathcal{O}(M^2,Ma^2,M^3).
\end{equation}
However, as the integration of \eqref{eq:intk} in the angular variable will produce a term proportional to 
\begin{equation}
\begin{aligned}
M\cos\varphi_R =M\cos\bigg(&\pi+a_1\f{M}{b}+a_2\f{Ma}{b}+a_3\f{M^2}{b^2}\\
&+\mathcal{O}(\f{Ma^2}{b^2},\f{M^3}{b^3})\bigg),
\end{aligned}
\end{equation}
with $a_1, a_2, a_3$ constants, it follows that after re-expanding this term in powers of $M$, only the first term in $M$ of $\varphi_R$ (proportional to $a_1$) will contribute at the desired third order.  
 Hence, in order to compute the deflection angle up to third order, we need to know the orbit a second order and the deflection angle at first order. Using the same argument it can be shown that in order to know the deflection angle at $n$ order using \eqref{eq:asa}, we need to know the orbit at $n-1$ order and the deflection angle at $n-2$.

To finalize, we need also to compute the contribution $\alpha_{\kappa_g}$ to the deflection angle. In our case \begin{equation}
\bm{\hat\beta}=-\frac{2Mar}{r^2-2 M r}d\varphi.
\end{equation}
The associated geodesic curvature is given by

\begin{equation}
\begin{split}
    \kappa_g=&-\frac{2aM}{\sqrt{r(r-2M)}[2M r(1-v^2)+r^2v^2]}\bigg|_{r_\gamma}.
    \end{split}
\end{equation}
%}

On the other hand, $dl$ is obtained from
\begin{equation}
    dl=\sqrt{\bigg(\hat{g}_{rr}(\frac{dr}{d\varphi})^2+\hat{g}_{\varphi\varphi}\bigg)\bigg|_{r_\gamma}}d\varphi.
\end{equation}
At the considered order we obtain that $\kappa_g dl$ is given by
\begin{equation}
\begin{split}
    \kappa_g dl\bigg|_{r_\gamma}=&\,d\varphi\bigg[-\frac{\sin\varphi}{v}\frac{2Ma}{b^2}
    \bigg.\\
    &\bigg.+\frac{2(\cos\varphi-1)(2v^2\cos\varphi+3v^2+1)}{v^3}\frac{M^2a}{b^3}\bigg].
    \end{split}
\end{equation}
Hence, the contribution $\alpha_{\kappa_g}$to $\alpha$ is obtained from
\begin{equation}\label{eq:alphakap}
\begin{split}
    \alpha_{\kappa_g}=&\int^{\varphi_R}_0 \kappa_g dl\approx \bigg[-\frac{4Ma}{b^2v}-\frac{2\pi(1+2v^2)}{v^3}\frac{M^2a}{b^3}\bigg].
    \end{split}
\end{equation}
Taking cognizance of \eqref{eq:alphaK} and \eqref{eq:alphakap}; and by using the adimensional parameter $\hat{a}=\frac{a}{M}$ (which for a black holes must satisfy $|\hat{a}|\leq 1$) the resulting deflection angle reads
\begin{widetext}
\begin{equation}\label{eq:alphaho}
\begin{split}
\alpha=&\underbrace{\frac{2M}{b}\left(1+\frac{1}{v^2}\right)}_{\alpha^{(1)}}+\underbrace{\bigg[\frac{3\pi}{4}\left(1+\frac{4}{v^2}\right)- \frac{4\hat{a}}{v}\bigg]\frac{M^2}{b^2}}_{\alpha^{(2)}}+\underbrace{\bigg[\frac{2}{3}\left(5+\frac{45}{v^2}+\frac{15}{v^4}-\frac{1}{v^6}\right)-\frac{2\pi(2+3v^2)\hat{a}}{v^3}+\frac{2(v^2+1)\hat{a}^2}{v^2}\bigg]\frac{M^3}{b^3}}_{\alpha^{(3)}}.
\end{split}
\end{equation}
\end{widetext}

\begin{figure}[h]
\centering
\includegraphics[clip,width=86mm]{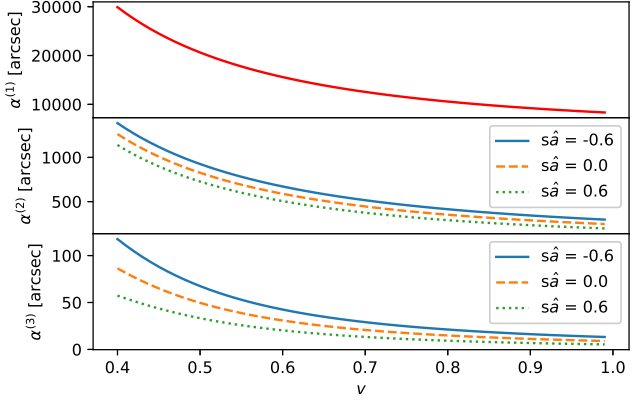}
\caption{Different contributions to the total deflection angle Eqn.\eqref{eq:alphaho} for a lens with parameters $M=4.1\times 10^6M_\odot$ and $\hat{a}=0.6$ in terms of the speed $v$ of the test massive particle.  We assume that $b=100M$.}
\label{leading-orderalphaconst}
\end{figure}

\begin{figure}[h]
\centering
\includegraphics[clip,width=86mm]{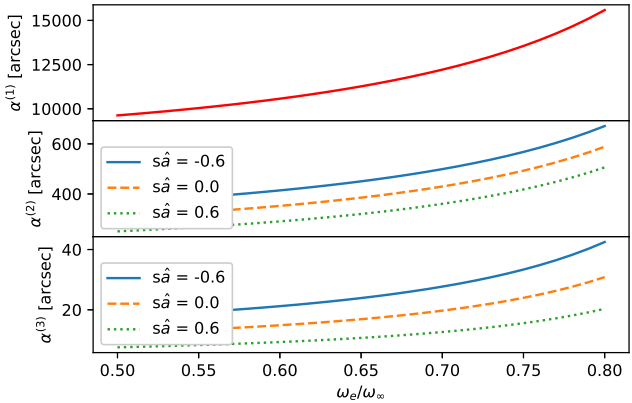}
\caption{Different contributions to the total deflection angle for a homogeneous plasma with $\omega_e=\text{constant}$ (obtained from Eqn.\eqref{eq:alphaho} by replacing $v^2$ by $1-\omega^2_e/\omega^2_\infty$). We consider a lens with parameters $M=4.1\times 10^6M_\odot$, $\hat{a}=0.6$ and $b=100M$.}
\label{higher-orderalphaconst}
\end{figure}

For massless particles it reduces to: 
\begin{equation}\label{eq:alphaholight}
\alpha=\frac{4M}{b}+\bigg[\frac{15\pi}{4}- 4\hat{a}\bigg]\frac{M^2}{b^2}+\bigg[\frac{128}{3}- 10\pi\hat{a}+4\hat{a}^2\bigg]\frac{M^3}{b^3};
\end{equation}
which agrees with known expressions  \cite{Iyer:2009hq,Aazami:2011tw}. As in the previous cases, for retrograde orbits we must to change  the sign of the terms  which are linear in $\hat{a}$.
Therefore, we have shown that the method proposed by Ono {\it et.al.}\cite{Ono:2017pie} can be applied in a successfully way to compute the deflection angle to higher order for stationary spacetimes, and moreover, it can be applied to massive particles.  We are not aware of a previous presentation of \eqref{eq:alphaho} in the literature. 

Note also that using the correspondence between the motion of test massive particles and photons in an homogeneous plasma, \eqref{eq:alphaho} also gives the deflection angle of light rays with frequency $\omega_\infty$ in a homogeneous plasma characterized by $\omega_e=\text{constant}$, by simply replacing $v$ in \eqref{eq:alphaho} by the group velocity $v_\text{gr}$ of the light ray in plasma given by $v_\text{gr}=n_0=\sqrt{1-\omega^2_e/\omega^2_\infty}$ in units where $c=1$. 
In Fig.\eqref{leading-orderalphaconst} and Fig.\eqref{higher-orderalphaconst} we consider a black hole with parameters similar to the super-massive black hole in the center of our galaxy Sgr A*, namely we take $M=4.1\times 10^6M_\odot$, $\hat{a}=0.6$ and we assume that $b=100M$. In both figures, we plot the different contributions $\alpha^{(1)},\alpha^{(2)}$ and $\alpha^{(3)}$ to the total deflection angle. In \eqref{leading-orderalphaconst} we plot the $\alpha^{(i)}$'s components in terms of the velocity of the massive particle and in Fig.\eqref{higher-orderalphaconst} in terms of the quotient $\omega_e/\omega_\infty$. For both situation we consider prograde and retrograde orbits ($s=\pm 1$) and also the nonspinning case $\hat{a}=0$.

In Appendix \eqref{Ap:B} we will present and alternative derivation of \eqref{eq:alphaho} using a method originally proposed by Aazami, Keeton and Petters \cite{Aazami:2011tw}.

\subsection{Deflection angle of light rays in a non-homogeneous plasma up to third order in a Kerr background}\label{third-order2}
Let us consider a stationary and axysymmetric spacetime surrounded by a cold nonmagnetized plasma and whose restriction to the equatorial plane has the form \eqref{eq:equati}.
In such a situation, the orbit equation can be derived to the Hamiltonian \eqref{ham-1} resulting in
\begin{equation}
    \bigg(\f{du}{d\varphi}\bigg)^2=-\f{u^4(\omega^2_e\Delta-Dp^2_t-2Hp_tp_\varphi+Ap^2_\varphi)\Delta}{B(-Hp_t+Ap_\varphi)^2}
\end{equation}
In general for a prograde photon propagating in a plasma environment $p_t=-\hbar \omega_\infty$ and $p_\varphi=-p_t b n_0$, with $b$ the impact parameter and $n_0$ the asymptotic value of the refraction index. In the next discussion we assume that the electronic number density profile is given by,
\begin{equation}\label{eq:plasmadensityprofile1}
\omega_{e0}=0, \ \ \  N_1(r)=\frac{\tilde{N}_0}{r^k},
\end{equation}
with $\tilde{N}_0$ a given constant {and $k$ a positive integer number. In the case of spherically symmetric spacetimes a detailed discussion of the leading order contribution to the deflection angle and also to associated optical quantities produced by this kind of density profiles was carried out by Bisnovatyi-Kogan and Tsupko in\cite{BisnovatyiKogan:2010ar}. In particular, this family of profiles can be useful in the study of plasma environments in galaxies and galaxy-clusters. Note also, that for the case of the supermassive black hole in the center of our galaxy Sgr A*, a plasma density profile with a radial dependence of the form $r^{-1.1}$ has also been considered by different authors\cite{Psaltis_2012,10.1093/mnras/stz138}. Even when it is not exactly in the family described by Eqn.\eqref{eq:plasmadensityprofile1}, for $k=1$ we can have an estimation of the behaviour of the deflection angle for this kind of system. Higher powers of $k$ (including $k=6$ and $k=16$) also appear in the discussion of light ray propagation in an extended solar corona model\cite{Tyler-1977,giampieri1996,Bertotti-1998,Turyshev_2019,Crisnejo-finite}, } { In \cite{BisnovatyiKogan:2010ar}  analytical closed expressions for the deflection angle were derived for any $k$. To higher order, a general expression is more difficult to obtain, and the intermediate steps are not very illustrative. For this reason, in the following we only consider the cases $k=\{1,2,3\}$, showing in detail the way to proceed for the case $k=2$, and only given the final expressions for the other two cases. For any other value of $k$ not considered in this work, the expression for the deflection angle can be found in exactly the same way as in the case $k=2$. Note also that our final expressions can not only be used when the rotation of the spacetime is taken into account, they can also be useful for spherically symmetric spacetimes if higher order corrections due to the presence of plasma need to be considered. In the following,} we also assume that the frequency $\omega_\infty$ of the photon at the asymptotic region is much bigger than the value of plasma electronic frequency $\tilde{\omega}_e\equiv \omega_e(b)$ at $r=b$. That is, we assume that $\epsilon=\f{\tilde{\omega}^2_e}{\omega^2_\infty}\ll 1$. Note that for the present situation, $n_0=1$.
Hence, the orbit equation can be rewritten as
\begin{equation}
    \bigg(\f{du}{d\varphi}\bigg)^2=-\f{u^4[\omega^2_e\Delta-(D+2Hb-Ab^2)\omega^2_\infty]\Delta}{B\omega^2_\infty(H+Ab)^2}
\end{equation}
Let us restrict our attention to the Kerr metric with the functions $A$, $H$, $B$ and $D$ given by \eqref{ew:a1}-\eqref{ew:a4}.
Using the ansatz,
\begin{equation}
\begin{split}
u=&u_0(\varphi)+u_{M}(\varphi)\gamma+u_a(\varphi)\delta+u_\epsilon(\varphi)\epsilon +u_{Ma}(\varphi)\gamma\delta\\
&+u_{\epsilon\,a}(\varphi)\epsilon\,a+u_{M\epsilon}(\varphi)\gamma\epsilon+u_{M}(\varphi)\gamma^2+u_{a^2}(\varphi)\delta^2\\ &+u_{\epsilon^2}(\varphi)\epsilon^2,
\end{split}
\end{equation}
 and by imposing the asymptotic condition that $\lim_{\varphi\to 0}u=0$, result in the following iterative solution for $u_\gamma$:
\begin{eqnarray}
%\begin{aligned}
u_0&=&\frac{\sin\varphi}{b},\\
u_m&=&\f{(\cos\varphi-1)^2}{b},\\
u_a&=&u_{\epsilon\,a}=0,\\
u_{\epsilon}&=&-\f{\cos\varphi(\tan\varphi-\varphi)}{2b}\\
u_{Ma}&=&-\f{2(1-\cos\varphi)}{b}\\
u_{M\epsilon}&=&-\f{(\cos\varphi-1)(\cos\varphi+\varphi\sin\varphi-1)}{b},\\
u_{M^2}&=&-\frac{1}{4b}[3\sin\varphi\cos^2\varphi+15\varphi\cos\varphi\\\nonumber &&-2\sin\varphi(1+8\cos\varphi)],\\
u_{a^2}&=&\f{\sin^3\varphi}{b},\\
u_{\epsilon^2}&=&-\f{[(\varphi^2-3)\sin\varphi+3\varphi\cos\varphi]}{8b}
\end{eqnarray}
At the considerer order $\mathcal{K}dS$ reads;
\begin{equation}
\begin{split}
    \mathcal{K}dS=&\bigg[\underbrace{2M}_{\mathcal{\hat{K}}_M}\,\,\underbrace{-2b^2\epsilon\,u}_{\mathcal{\hat{K}}_{\epsilon}}\,\,\,\underbrace{+3M^2u}_{\mathcal{\hat{K}}_{M^2}}\,\,\underbrace{+12Mb^2\epsilon\,u^2}_{\mathcal{\hat{K}}_{M\epsilon}}\,\,\underbrace{-4b^4\epsilon^2 u^3}_{\mathcal{\hat{K}}_{\epsilon^2}}\bigg.\\
    &\bigg.\,\,\underbrace{+6M^3 u^2}_{\mathcal{\hat{K}}_{M^3}}\,\,\underbrace{-10b^2\epsilon\,M^2 u^3}_{\mathcal{\hat{K}}_{\epsilon\,M^2}}\,\,\underbrace{+6Ma^2u^2 }_{\mathcal{\hat{K}}_{Ma^2}}\,\,\underbrace{+30 M b^4\epsilon^2 u^4}_{\mathcal{\hat{K}}_{M\epsilon^2}}\,\,\bigg.\\
    &\bigg.\underbrace{-4b^2\epsilon\,a^2 u^3}_{\mathcal{\hat{K}}_{\epsilon\,a^2}}\,\,\underbrace{-6b^6\epsilon^3 u^5}_{\mathcal{\hat{K}}_{\epsilon^3}}\,\,\bigg]dud\varphi.
    \end{split}
\end{equation}

In order to compute the deflection angle to higher order, we also need to know the expression for the deflection angle at first order. It can be directly obtained from Eqn.\eqref{eq:signoi} with the result: 
\begin{equation}\label{eq:117}
    \alpha^{(1)}=\frac{4M}{b}-\frac{\pi\epsilon}{2}-\frac{4Ma}{b^2}.
%\end{equation}
\end{equation}
Therefore the contribution $\alpha_{\mathcal{K}}$ to the deflection angle including third order terms is determined by\footnote{As in Section \eqref{third-order}, it can be checked that adding terms to $\varphi_R$ with corrections of order 2 in $\gamma, \delta$ and $\epsilon$ (note that the last term in Eqn.\eqref{eq:117} is one of them) do not contribute to the deflection angle at the considered order. More precisely, the only terms that could potentially produce corrections by adding these extra terms come from the integration of $\tilde{\mathcal{K}}_M$ and $\tilde{\mathcal{K}}_\epsilon$. 
However using similar arguments as in Section \eqref{third-order} it can be shown that the only terms that contribute at the considered order, are in fact the first two terms present in $\alpha^{(1)}$.}
\begin{equation}\label{eq:alphaK2}
\begin{split}
    \alpha_{\mathcal{K}}=&\int^{\varphi_R=\pi+\alpha^{(1)}}_0\int^{u_\gamma}_0 \mathcal{K}dS\\
    =&\f{4M}{b}-\epsilon\f{\pi}{2}+\f{15\pi\,M^2}{4b^2}-\f{4M\epsilon}{b}\\
    &+\f{3}{8}\pi\epsilon^2+\f{128M^3}{3b^3}-\f{4\pi\,a{M^2}}{b^3}-\f{45\pi\epsilon\,M^2}{8b^2}\\
    &+\f{4a^2M}{b^3}+\f{8Ma\epsilon}{b^2}+\f{4M\epsilon^2}{b}-\f{3\pi\epsilon\,a^2}{4b^2}-\f{5\pi}{16}\epsilon^3.
    \end{split}
\end{equation}
The geodesic curvature reads,
\begin{equation}
    \kappa_g=-\f{2M a u^3}{[1-\epsilon\,b^2 u^2(1-2M u)]\sqrt{1-2M u}}\bigg|_{u_\gamma}
\end{equation}
while $dl$ is given by
\begin{equation}
\begin{split}
    dl=&\sqrt{\bigg(\hat{g}_{rr}(\frac{dr}{d\varphi})^2+\hat{g}_{\varphi\varphi}\bigg)\bigg|_{r_\gamma}}d\varphi\\
    =&\sqrt{\bigg(\hat{g}_{rr}\bigg(\f{1}{u^2}\frac{du}{d\varphi}\bigg)^2+\hat{g}_{\varphi\varphi}\bigg)\bigg|_{u_\gamma}}d\varphi\\
    =&\f{[1-\epsilon\,b^2u^2(1-2M u)](1+a^2u^2-2M u)}{u^2[2aM u+b(1-2M u)](1-2M u)}\bigg|_{u_\gamma}d\varphi.
    \end{split}
\end{equation}
Hence, 
\begin{equation}
\begin{split}
    \kappa_g dl=&-\f{2M a u(1+a^2u^2-2M u)}{(1-2M u)^{3/2}[2a M u+b(1-2M u)]}\bigg|_{u_\gamma}d\varphi\\
    \approx & \bigg[-\f{2M a}{b^2}\sin\varphi+\f{4aM^2}{b^3}(\cos\varphi+2)(\cos\varphi-1)\bigg.\\
    &\bigg.+\cos\varphi(\tan\varphi-\varphi)\f{aM\epsilon}{b^2}\bigg]d\varphi
    \end{split}
\end{equation}
Therefore, the contribution $\alpha_{\kappa_g}$ to $\alpha$ is 
\begin{equation}\label{eq:alphakap2e}
\begin{split}
    \alpha_{\kappa_g}=&\int^{\varphi_R}_0 \kappa_g dl\approx\bigg[-4\f{aM}{b^2}-\f{6\pi\,aM^2}{b^3}+\f{4aM\epsilon}{b^2}\bigg].
    \end{split}
\end{equation}

Finally, collecting the results of eqs.\eqref{eq:alphaK2} and \eqref{eq:alphakap2e}, and writing them in terms of $\hat{a}=a/M$
we obtain:
\begin{equation}
    \alpha=\alpha_{\text{vac}}+\alpha_\text{p}
\end{equation}
with
\begin{equation}\label{eq:alphaplaspi}
\begin{split}
    \alpha_\text{p}\approx& \,\,-\epsilon\f{\pi}{2}+\f{3}{8}\pi\epsilon^2-\f{5}{16}\pi\epsilon^3-\f{4M\epsilon}{b}+\f{4M\epsilon^2}{b}\\
    &-\bigg[\f{45}{2}\pi
    -48\hat{a}+3\pi\,\hat{a}^2\bigg]\f{\epsilon\,M^2}{4b^2},
    \end{split}
\end{equation}
the contribution due to the presence of the plasma
and $\alpha_{\text{vac}}$ the value of the deflection angle for light rays propagating in vacuum as given by Eqn.\eqref{eq:alphaholight}.

\begin{figure}[h]
\centering
\includegraphics[clip,width=86mm]{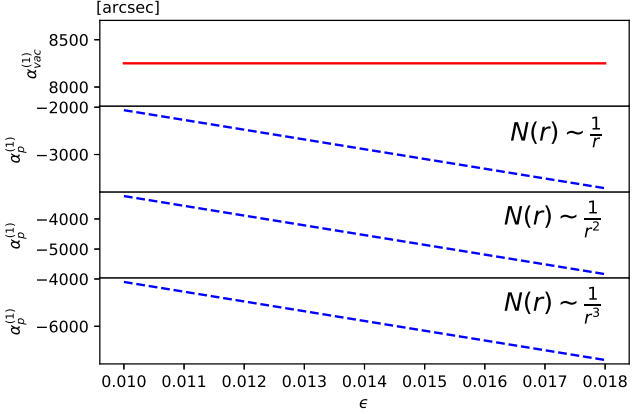}
\caption{Leading order contribution to the deflection angle for a lens with parameters $M=4.1\times 10^6M_\odot$ and $\hat{a}=0.6$ and different plasma density profiles of the form $N(r)=\tilde{N}_0r^{-k}$. We assume that $b=100M$. { Here $\alpha^{(1)}_\text{vac}=4M/b$ and $\alpha^{(1)}_\text{p}=-\epsilon$ for $k=1$, $\alpha^{(1)}_\text{p}=-\epsilon \pi/2$ for $k=2$ and $\alpha^{(1)}_\text{p}=-2\epsilon$ for $k=3$. All of them are particular cases of the expression \eqref{eq:exactgammaexp}. Here $\epsilon\in [0.01,0.018]$.}}
\label{leading-order}
\end{figure}

\begin{figure}[h]
\centering
\includegraphics[clip,width=86mm]{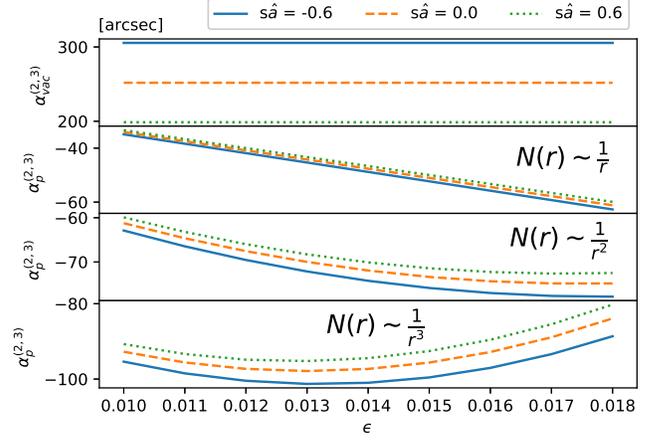}
\caption{Higher order contribution to the deflection angle for a lens with parameters $M=4.1\times 10^6M_\odot$ and $\hat{a}=0.6$. We assume that $b=100M$. Here $\alpha^{(2,3)}_{\text{vac}}=\alpha_{\text{vac}}-\alpha^{(1)}_{\text{vac}}$ 
and $\alpha^{(2,3)}_{\text{p}}=\alpha_{\text{p}}-\alpha^{(1)}_{\text{p}}$. $\epsilon\in [0.01,0.018]$.  As in Fig.\eqref{leading-order} each of them is considered for the different density profiles determined by $k=1,2$ and 3. }
\label{higer-order}
\end{figure}
The first three terms in Eqn.\eqref{eq:alphaplaspi} represent the contribution to the deflection angle due to pure plasma and the rest of the terms give the contribution to the deflection angle due to the coupling between plasma and gravitational effects. As before, the deflection angle for retrograde orbits is obtained from \eqref{eq:alphaplaspi} by changing the sign in all the terms linear in $\hat{a}.$

The contribution of each component to the deflection angle will depend on the parameters $M,a,b$ and the quotient $\epsilon$ between the electronic frequency of the plasma and the photon frequency $\omega_\infty.$ 

{Now, we write without derivation expressions for the deflection angle in the cases $k=1$ and $k=3$ obtained using similar steps. For $k=1$, i.e. $N=\tilde{N}_0/r$, the final expression for the deflection angle reads:
\begin{equation}
    \alpha=\alpha_{\text{vac}}+\alpha_{\text{p}},
\end{equation}
with $\alpha_{\text{p}}$ given by
\begin{equation}
    \alpha_{\text{p}}=-\epsilon+\f{1}{12}\epsilon^3-\f{\pi M\epsilon}{2b}+(2\pi\hat{a}-\hat{a}^2-8)\f{\epsilon M^2}{b^2}.
\end{equation}
Note that for this case, there is not a pure plasma contribution at second order in $\epsilon$. 
Similarly, for $k=3$, i.e. $N=\tilde{N}_0/r^3$, the  contribution $\alpha_{\text{p}}$ to the deflection angle reads:
\begin{equation}
    \alpha_{\text{p}}=-2\epsilon+\f{15\pi}{16}\epsilon^2-\f{16}{3}\epsilon^3-\f{9\pi M\epsilon}{4b}+(6\pi\hat{a}-4\hat{a}^2-32)\f{\epsilon M^2}{b^2}.
    \end{equation}
In all the cases, the linear term in $\epsilon$ can be considered as particular cases of the known expression  obtained in the past by Giampieri\cite{giampieri1996} and reobtained in a more general context by Bisnovatyi-Kogan and Tsupko in \cite{BisnovatyiKogan:2010ar}:
\begin{equation}\label{eq:exactgammaexp}
    \alpha^{(1)}_\text{p}=-\epsilon\sqrt{\pi}\f{\Gamma(\f{k+1}{2})}{\Gamma(\f{k}{2})}
\end{equation}
with $\Gamma(x)$ the Gamma function.  }

In Fig.\eqref{leading-order} we consider a black hole with the same parameters as in Fig.\eqref{leading-orderalphaconst} and Fig.\eqref{higher-orderalphaconst} assuming again that $b=100M$. We plot the leading order contribution to the deflection angle given by $\alpha^{(1)}_\text{vac}=4M/b$ and $\alpha^{(1)}_\text{p}$ for the considered cases in terms of $\epsilon\in [0.01,0.018]$. For this range of values, the deflection angle produced by the plasma is of the same order as the Kerr vacuum angle (but with opposite direction). {That is, as is well known, and can also be directly seen from Eqn.\eqref{eq:signoi} the effect of a non uniform plasma that decays with $r$ is to produce a divergent lens effect. Note that for the same $\epsilon$ value, the faster the plasma profile decays, the greater the divergence effect on the light rays.}  Of course, it is a chromatic effect, and different values of $\epsilon$ will contribute at different ways.
Similarly, in Fig.\eqref{higer-order} we plot the contribution to the deflection angle due to the rest of the terms in $\alpha_\text{vac}$ and $\alpha_\text{p}$ {for the three considered plasma density profiles.} The case $s=\pm 1$ corresponding to prograde/retrograde orbits respectively. For completeness, we also plot the contributions to the deflection angle for a nonrotating BH ($\hat{a}=0$).

\section{Conclusion}\label{sec-7}
In this paper we have used the Gauss-Bonnet theorem to study the gravitational deflection of massive particles and light rays in stationary spacetimes. For that purpose we have obtained a Finsler-Randers type metric which was utilized to calculate the deflection angle of light rays and massive particles. More specifically, using the Finsler-Randers metric for stationary spacetimes along with the correspondence between the motion of light rays in a plasma medium and relativistic test massive particles we have extended the recent method proposed by Werner. As particular examples we have obtained the deflection angle in a rotating black hole and a rotating wormhole geometry. 
However, the most important result arising from our study, is the fact that we have been able to obtain higher order corrections in the expression for the deflection angle in the Kerr spacetime background applying the Ono {\it et.al.} approach. We subsequently analysed also the deflection angle for test massive particles and light rays propagating in a non homogeneous cold plasma by including third order correction terms. 
\section*{Acknowledgments}

G.C. is a CNPq-TWAS fellow. E.G. acknowledges support from CONICET and SeCyT-UNC. The authors thank
the anonymous referee for the valuable suggestions. 

 \appendix

\section{Rotating Teo wormhole trought the Werner method}\label{app}

For completeness, we apply the Werner method to the rotating Teo solution given by the Eqn.\eqref{Teo-1}.
In the equatorial plane $\theta=\pi/2$, for the linearized case we find the corresponding Kerr-Randers optical metric given by
\begin{eqnarray}\notag
&\mathcal{F}&\left(r,\varphi,\frac{\mathrm{d}r}{\mathrm{d}t},\frac{\mathrm{d}\varphi}{\mathrm{d}t}\right)=\left[1-(1-v^2)\right]^{1/2}\\
&\times& \left[r^2 \left(\frac{\mathrm{d}\varphi}{\mathrm{d}t}\right)^2+\frac{r}{r-b_0}\left(\frac{\mathrm{d}r}{\mathrm{d}t}\right)^2\right]^{1/2}-\frac{2a}{r}\frac{\mathrm{d}\varphi}{\mathrm{d}t}.
\end{eqnarray}

In a similar way as in the Kerr metric, we can construct a Riemannian manifold $(\mathcal{M},\bar{g})$ which osculates the Randers manifold using a vector field $\bar{X}$ tangent to the geodesic $\gamma_{\mathcal{F}}$, such that $\bar{X}(\gamma_{\mathcal{F}})=\dot{x}$.
Finally going through the same procedure as in the case of Kerr black hole, for the metric components of the osculating Riemannian geometry we find the following equations
\begin{widetext}
\begin{eqnarray}
\bar{g}_{rr}&=&\frac{v^2 r}{r-r_0}-\frac{ 2 \sin^6\varphi r^2 a v }{(r-r_0) b^3 \left[ \frac{r \cos^2 \varphi}{r-b_0}+\frac{r^2 \sin^4\varphi}{b^2}\right]^{3/2}}+\mathcal{O}({\frac{a^2}{b^2}}),\\
\bar{g}_{\varphi \varphi}&=&r^2 v^2 -\frac{6  r^2 v a \sin^2 \varphi \left( 2 r(r-b_0)\sin^4\varphi+3 b^2 \cos^2\varphi  \right)}{3 (r-r_0) b^3 \left[ \frac{r \cos^2 \varphi}{r-b_0}+\frac{r^2 \sin^4\varphi}{b^2}\right]^{3/2}  }+\mathcal{O}({\frac{a^2}{b^2}}),\\ 
\bar{g}_{r\varphi}&=& \frac{2  v r a \cos^3\varphi}{(r-b_0)^2\left[ \frac{r \cos^2 \varphi}{r-b_0}+\frac{r^2 \sin^4\varphi}{b^2}\right]^{3/2} }+\mathcal{O}({\frac{a^2}{b^2}}),
\end{eqnarray}
with the determinant given as
\begin{equation}
\det \bar{g}= \frac{r^3 v^4}{r-b_0}+\frac{6 a r^3 v^3 \sin^2\varphi \left(r \sin^4\varphi b_0-r^2 \sin^4\varphi -\cos^2\varphi b^2   \right)}{(r-b_0)^2 b^3 \left[ \frac{r \cos^2 \varphi}{r-b_0}+\frac{r^2 \sin^4\varphi}{b^2}\right]^{3/2}}+\mathcal{O}(a^2).
\end{equation}

Using the above relations we find the following result for the Gaussian optical curvature
\begin{eqnarray}
\mathcal{K}&=& -\frac{b_0}{2 v^2 r^3}+\frac{3 a \sin^2\varphi}{r^3 v^8}f(r,\varphi)+\mathcal{O}({\frac{a^2}{b^4}})
\end{eqnarray}
where
\begin{eqnarray}\notag
f(r,\varphi)&=& \frac{1}{\left[ \frac{r^2 \sin^4 \varphi +b^2 \cos^2\varphi}{b^2}  \right]^{5/2}\left[r^2 \sin^4 \varphi +b^2 \cos^2\varphi  \right]^{7/2}}\Big[\Big(-2 \sin^8\varphi b^2 r^4-3 b^4 r^2 \cos^2\varphi \sin^4\varphi-\cos^4\varphi b^6\Big)\\\notag
&\times& v^7 \left[\frac{r^2 \sin^4 \varphi}{b^2} +\cos^2\varphi  \right]^{7/2}+\Big(4 r^6 \sin^{12}\varphi-b^2 r^4 \sin^{10}\varphi+b^2 r^4 \cos^2\varphi \sin^8\varphi+8 b^3 r^3 \cos^2\varphi \sin^7\varphi\\\notag
&+& \left(-10 b^2 r^4 \cos^4\varphi+4 b^4 r^2 \cos^2\varphi\right) \sin^6\varphi+16 b^3 r^3 \cos^4\varphi \sin^5\varphi+13 b^4 r^2 \cos^4\varphi \sin^4\varphi-2 b^5 r\cos^6\varphi \sin^2\varphi\\
&+& 10b^4 r^2 \cos^6\varphi \sin^2\varphi-4 b^5 r \cos^6\varphi+b^6 \cos^6\varphi \Big)v^7 \left[\frac{r^2 \sin^4 \varphi +b^2 \cos^2\varphi}{b^2}  \right]^{5/2}\Big].
\end{eqnarray}
\end{widetext}
From \eqref{alpha-werner} the deflection angle can is given by,
\begin{equation}
\alpha_{\text{mp}} = - \int\limits_{0}^{\pi}\int\limits_{\frac{b}{\sin \varphi}}^{\infty}\left(-\frac{b_0}{2 v r^2}+\frac{3 a \sin^2\varphi}{r^2 v^{8}}f(r,\varphi)\right) \mathrm{d}r\,\mathrm{d}\varphi.
\end{equation}
Evaluating this expression the deflection angle results
\begin{equation} \label{alp2}
\alpha_{\text{mp}} = \frac{b_0}{b}- \frac{ 4a}{b^2 v},
\end{equation}
which completely agrees with  \eqref{eq:teoangle}.

\section{Deflection of light rays up to the third order in a Kerr spacetime using direct integration}\label{Ap:B}
As an independent check of our expression for the deflection angle given by \eqref{eq:alphaho} we will use a direct integration of the orbit equation following a straightforward extension of the method presented by Aazami, Keeton and Petters and Hansen in \cite{Aazami:2011tw} orignally introduced to study null geodesics. 

The orbit equation for both prograde ($s=+1$) and retrograde orbits ($s=-1$) in the equatorial plane can be reexpressed as:
\begin{widetext}
\begin{equation}\label{eq:orbitagener}
    \f{d\varphi}{dr}=\pm\f{s\sqrt{r}b(r-2M F)}{[a^2+r(r-2M)]\sqrt{r^3+b^2(2M F^2-Gr)+2M \Gamma\,r^2}},
\end{equation}
\end{widetext}
where
\begin{eqnarray}
F&\equiv&1-s\f{a}{bv}=1-s\hat{a}\f{M}{bv},\\
G&\equiv&1-\f{a^2}{b^2}=1-\hat{a}^2\f{M^2}{b^2},\\
\Gamma&=&\f{1-v^2}{v^2}.
\end{eqnarray}
Expression \eqref{eq:orbitagener} reduces to Eqn.(B14) of \cite{Aazami:2011tw} in the case of null rays ($\Gamma=0$).
The deflection angle is obtained as in the usual way by
\begin{equation}\label{eq:alphahan}
    \alpha=2\int^\infty_{r_0}\bigg|\f{d\varphi}{dr}\bigg|dr-\pi,
\end{equation}
where $r_0$ the value of the radial coordinate at the closest approach to the lens.
The relation between the impact parameter $b$ and $r_0$ follows from require $\f{dr}{d\varphi}\bigg|_{r=r_0}=0$, which from \eqref{eq:orbitagener} implies that
\begin{equation}
    r^3_0+b^2(2M F^2-Gr_0)+2m\Gamma\,r^2_0=0.
\end{equation}
The positive real solution of the last equation is 
\begin{equation}\label{eq:busa}
    b=\f{\sqrt{r_0+2M\Gamma}}{\sqrt{G-\f{2M}{r_0}F^2}}.
\end{equation}
By substituting Eqn.\eqref{eq:busa} into \eqref{eq:orbitagener}, and changing the integration variable $r$ in Eqn.\eqref{eq:alphahan} by $x=\f{r_0}{r}$, the deflection angle is determined by the following expression:
\begin{widetext}
\begin{equation}\label{eq:alphafhp}
    \alpha=2\int^1_0\f{\sqrt{1+2\Gamma h}(1-2hFx)}{(1-2hx+\hat{a}^2h^2x^2)\bigg\{G(1-x^2)-2F^2h(1-x^3)+\Gamma(1-x)h[2Gx-4x(1+x)F^2h]\bigg\}^{1/2}}dx-\pi;
\end{equation}
\end{widetext}
with $h=M/r_0$.

By assuming that $h\ll 1$, we can do a Taylor expansion in $h$ of the integrand of \eqref{eq:alphafhp}, which after integration and  conserving terms up to third order in $h$ gives,
\begin{equation}\label{eq:alphasta}
    \alpha=b_0\pi+a_1h+(a_2+b_2\pi)h^2+(a_3+b_3\pi)h^3+\mathcal{O}(h^4);
\end{equation}
with
\begin{widetext}
\begin{eqnarray}
b_0&=&(\f{1}{G^{1/2}}-1),\\
a_1&=&\f{2\{2[G(1-F)+F^2]+\Gamma\,G\}}{G^{3/2}},\\
a_2&=&-a_1\f{F^2+\Gamma\,G}{G},\\
b_2&=&\f{15F^4-4G(F-1)(3F^2+2G)-2\hat{a}^2G^2+4\Gamma\,G[3F^2+2G(1-F)]}{4G^{5/2}},\\
a_3&=&\f{1}{3G^{7/2}}\bigg\{122F^6-90F^4(F-1)G-16F^2(-4+\hat{a}^2+4F)G^2+8(4+\hat{a}^2(F-2)-4F)G^3\bigg.\nonumber\\
&&\bigg.+\Gamma[12G^2(1-F)(9F^2+4G)+153F^4G-12\hat{a}^2G^3]+\Gamma^2(30G^3(1-F)+48G^2F^2)+7\Gamma^3G^3\bigg\},\\
b_3&=&-2b_2\f{F^2+\Gamma\,G}{G}.
\end{eqnarray}
\end{widetext}
These expressions generalize the results found in \cite{Aazami:2011tw} to the case of test massive particles. We can also write the deflection angle in terms of an expansion in $\gamma=M/b$  using that $h$  can be perturbatively obtained from \eqref{eq:busa},
\begin{equation}
\begin{split}
   h=&\f{1}{\sqrt{G}}\gamma+\f{F^2+\Gamma\,G}{G^2}\gamma^2+\f{5F^4+\Gamma\,G(6F^2+\Gamma\,G)}{2G^{7/2}}\gamma^3\\
   &+\mathcal{O}(\gamma^4).
    \end{split}
\end{equation}
In terms of $\gamma$ the deflection angle reads:
\begin{equation}\label{eq:alffinalhp}
    \alpha=A_0\pi+A_1\gamma+A2\gamma^2+A_3\gamma^3+\mathcal{O}(\gamma^4),
\end{equation}
with
\begin{eqnarray}
    A_0&=&b_0,\\
    A_1&=&\f{a_1}{\sqrt{G}},\\
    A_2&=&\f{a_1(F^2+\Gamma\,G)+G(a_2+b_2\pi)}{G^2},\\
    A_3&=&\f{1}{2G^{7/2}}\bigg\{\bigg[6a_1\Gamma\,F^2+4F^2(a_2+b_2\pi)\bigg]G\bigg.\nonumber\\
    &&
   + \bigg[\Gamma(\Gamma\,a_1+4a_2+4b_2\pi)+2(a_3+b_3\pi)\bigg]G^2\bigg.\nonumber\\
    &&\bigg.+5a_1F^4\bigg\}
\end{eqnarray}

Of course, as $F$ and $G$ are already functions of $\gamma$, the $A_i$ coefficients must be expanded in $\gamma$, and after that, by replacing in Eqn.\eqref{eq:alffinalhp} we recover the deflection angle as given by Eqn.\eqref{eq:alphaho} with the linear terms in $\hat{a}$ multiplied by $s$.

%\bibliographystyle{unsrt}
%\bibliography{biblio}

\begin{thebibliography}{10}
	
	\bibitem{Crisnejo-gauss-bonnet-1}
	Gabriel Crisnejo and Emanuel Gallo.
	\newblock {Weak lensing in a plasma medium and gravitational deflection of
		massive particles using the Gauss-Bonnet theorem. A unified treatment}.
	\newblock {\em Phys. Rev.}, D97(12):124016, 2018.
	
	\bibitem{Akiyama:2019cqa}
	Kazunori Akiyama et~al.
	\newblock {First M87 Event Horizon Telescope Results. I. The Shadow of the
		Supermassive Black Hole}.
	\newblock {\em Astrophys. J.}, 875(1):L1, 2019.
	
	\bibitem{Akiyama:2019brx}
	Kazunori Akiyama et~al.
	\newblock {First M87 Event Horizon Telescope Results. II. Array and
		Instrumentation}.
	\newblock {\em Astrophys. J.}, 875(1):L2, 2019.
	
	\bibitem{Akiyama:2019sww}
	Kazunori Akiyama et~al.
	\newblock {First M87 Event Horizon Telescope Results. III. Data Processing and
		Calibration}.
	\newblock {\em Astrophys. J.}, 875(1):L3, 2019.
	
	\bibitem{Akiyama:2019bqs}
	Kazunori Akiyama et~al.
	\newblock {First M87 Event Horizon Telescope Results. IV. Imaging the Central
		Supermassive Black Hole}.
	\newblock {\em Astrophys. J.}, 875(1):L4, 2019.
	
	\bibitem{Akiyama:2019fyp}
	Kazunori Akiyama et~al.
	\newblock {First M87 Event Horizon Telescope Results. V. Physical Origin of the
		Asymmetric Ring}.
	\newblock {\em Astrophys. J.}, 875(1):L5, 2019.
	
	\bibitem{Akiyama:2019eap}
	Kazunori Akiyama et~al.
	\newblock {First M87 Event Horizon Telescope Results. VI. The Shadow and Mass
		of the Central Black Hole}.
	\newblock {\em Astrophys. J.}, 875(1):L6, 2019.
	
	\bibitem{Perlick-photon-region-1}
	Arne Grenzebach, Volker Perlick, and Claus L{\"a}mmerzahl.
	\newblock {Photon Regions and Shadows of Kerr-Newman-NUT Black Holes with a
		Cosmological Constant}.
	\newblock {\em Phys. Rev.}, D89(12):124004, 2014.
	
	\bibitem{Gibbons-gauss-bonnet}
	G.~W. Gibbons and M.~C. Werner.
	\newblock {Applications of the Gauss-Bonnet theorem to gravitational lensing}.
	\newblock {\em Class. Quant. Grav.}, 25:235009, 2008.
	
	\bibitem{Werner:2012rc}
	M.~C. Werner.
	\newblock {Gravitational lensing in the Kerr-Randers optical geometry}.
	\newblock {\em Gen. Rel. Grav.}, 44:3047--3057, 2012.
	
	\bibitem{Ishihara:2016vdc}
	Asahi Ishihara, Yusuke Suzuki, Toshiaki Ono, Takao Kitamura, and Hideki Asada.
	\newblock {Gravitational bending angle of light for finite distance and the
		Gauss-Bonnet theorem}.
	\newblock {\em Phys. Rev.}, D94(8):084015, 2016.
	
	\bibitem{Ono:2017pie}
	Toshiaki Ono, Asahi Ishihara, and Hideki Asada.
	\newblock {Gravitomagnetic bending angle of light with finite-distance
		corrections in stationary axisymmetric spacetimes}.
	\newblock {\em Phys. Rev.}, D96(10):104037, 2017.
	
	\bibitem{Ono:2018ybw}
	Toshiaki Ono, Asahi Ishihara, and Hideki Asada.
	\newblock {Deflection angle of light for an observer and source at finite
		distance from a rotating wormhole}.
	\newblock {\em Phys. Rev.}, D98(4):044047, 2018.
	
	\bibitem{Kulsrud-1992}
	Russell Kulsrud and Abraham Loeb.
	\newblock Dynamics and gravitational interaction of waves in nonuniform media.
	\newblock {\em Phys. Rev. D}, 45:525--531, Jan 1992.
	
	\bibitem{BisnovatyiKogan:2010ar}
	G.~S. Bisnovatyi-Kogan and O.~{\relax Yu}. Tsupko.
	\newblock {Gravitational lensing in a non-uniform plasma}.
	\newblock {\em Mon. Not. Roy. Astron. Soc.}, 404:1790--1800, 2010.
	
	\bibitem{Tsupko:2013cqa}
	Oleg~Yu Tsupko and Gennady~S. Bisnovatyi-Kogan.
	\newblock {Gravitational lensing in plasma: Relativistic images at homogeneous
		plasma}.
	\newblock {\em Phys. Rev.}, D87(12):124009, 2013.
	
	\bibitem{Tsupko-strong}
	Oleg~Yu. Tsupko.
	\newblock Unbound motion of massive particles in the schwarzschild metric:
	Analytical description in case of strong deflection.
	\newblock {\em Phys. Rev. D}, 89:084075, Apr 2014.
	
	\bibitem{Crisnejo-charged-particles}
	Gabriel Crisnejo, Emanuel Gallo, and Jos\'e~R. Villanueva.
	\newblock Gravitational lensing in dispersive media and deflection angle of
	charged massive particles in terms of curvature scalars and energy-momentum
	tensor.
	\newblock {\em Phys. Rev. D}, 100:044006, Aug 2019.
	
	\bibitem{Gallo-lens-2011}
	Emanuel Gallo and Osvaldo~M. Moreschi.
	\newblock Gravitational lens optical scalars in terms of energy-momentum
	distributions.
	\newblock {\em Phys. Rev. D}, 83:083007, 2011.
	
	\bibitem{Gallo-peculiar}
	Emanuel Gallo and Osvaldo Moreschi.
	\newblock {Peculiar anisotropic stationary spherically symmetric solution of
		Einstein equations}.
	\newblock {\em Mod. Phys. Lett.}, A27:1250044, 2012.
	
	\bibitem{Boero:2016nrd}
	Ezequiel~F Boero and Osvaldo~M Moreschi.
	\newblock Gravitational lens optical scalars in terms of energy-momentum
	distributions in the cosmological framework.
	\newblock {\em Mon. Not. Roy. Astron. Soc.}, 475(4):4683--4703, 2018.
	
	\bibitem{Crisnejo-lens-2018}
	Gabriel Crisnejo and Emanuel Gallo.
	\newblock Expressions for optical scalars and deflection angle at second order
	in terms of curvature scalars.
	\newblock {\em Phys. Rev. D}, 97:084010, 2018.
	
	\bibitem{Solar-Radio}
	D.~O. Muhleman and I.~D. Johnston.
	\newblock Radio propagation in the solar gravitational field.
	\newblock {\em Phys. Rev. Lett.}, 17:455--458, Aug 1966.
	
	\bibitem{Muhleman-1970}
	D.~O. Muhleman, R.~D. Ekers, and E.~B. Fomalont.
	\newblock Radio interferometric test of the general relativistic light bending
	near the sun.
	\newblock {\em Phys. Rev. Lett.}, 24:1377--1380, Jun 1970.
	
	\bibitem{Breuer-1980}
	R.~A. Breuer and J.~Ehlers.
	\newblock Propagation of high-frequency electromagnetic waves through a
	magnetized plasma in curved space-time. i.
	\newblock {\em Proceedings Mathematical Physical \& Engineering Sciences}, 370,
	03 1980.
	
	\bibitem{Breuer-1981a}
	R.~A. Breuer and J.~Ehlers.
	\newblock Propagation of high-frequency electromagnetic waves through a
	magnetized plasma in curved space-time. ii - application of the asymptotic
	approximation.
	\newblock {\em Proceedings of the Royal Society of London Series A}, 01 1981.
	
	\bibitem{Breuer-1981b}
	R.~A. Breuer and J.~Ehlers.
	\newblock Propagation of electromagnetic waves through magnetized plasmas in
	arbitrary gravitational fields.
	\newblock {\em Astronomy \& Astrophysics}, 03 1981.
	
	\bibitem{Perlick-book}
	Volker Perlick.
	\newblock {\em Ray Optics, Fermat{'}s Principle, and Applications to General
		Relatively}.
	\newblock Lecture Notes in Physics 61. Springer-Verlag Berlin Heidelberg, first
	edition, 2000.
	
	\bibitem{BisnovatyiKogan:2008yg}
	G.~S. Bisnovatyi-Kogan and O.~{\relax Yu}. Tsupko.
	\newblock {Gravitational radiospectrometer}.
	\newblock {\em Grav. Cosmol.}, 15:20--27, 2009.
	
	\bibitem{Tsupko:2014sca}
	O.~{\relax Yu}. Tsupko and G.~S. Bisnovatyi-Kogan.
	\newblock {Influence of Plasma on Relativistic Images of Gravitational
		Lensing}.
	\newblock {\em Nonlin. Phenom. Complex Syst.}, 17(4):455--457, 2014.
	
	\bibitem{Tsupko:2014lta}
	O.~{\relax Yu}. Tsupko and G.~S. Bisnovatyi-Kogan.
	\newblock {Gravitational lensing in the presence of plasmas and strong
		gravitational fields}.
	\newblock {\em Grav. Cosmol.}, 20(3):220--225, 2014.
	
	\bibitem{Perlick:2015vta}
	Volker Perlick, Oleg~{\relax Yu}. Tsupko, and Gennady~S. Bisnovatyi-Kogan.
	\newblock {Influence of a plasma on the shadow of a spherically symmetric black
		hole}.
	\newblock {\em Phys. Rev.}, D92(10):104031, 2015.
	
	\bibitem{Bisnovatyi-Kogan:2015dxa}
	G.~S. Bisnovatyi-Kogan and O.~{\relax Yu}. Tsupko.
	\newblock {Gravitational Lensing in Plasmic Medium}.
	\newblock 2015.
	\newblock [Plasma Phys. Rep.41,562(2015)].
	
	\bibitem{Perlick:2017fio}
	Volker Perlick and Oleg~{\relax Yu}. Tsupko.
	\newblock {Light propagation in a plasma on Kerr spacetime: Separation of the
		Hamilton-Jacobi equation and calculation of the shadow}.
	\newblock {\em Phys. Rev.}, D95(10):104003, 2017.
	
	\bibitem{Bisnovatyi-Kogan:2017kii}
	Gennady Bisnovatyi-Kogan and Oleg Tsupko.
	\newblock {Gravitational Lensing in Presence of Plasma: Strong Lens Systems,
		Black Hole Lensing and Shadow}.
	\newblock {\em Universe}, 3(3):57, 2017.
	
	\bibitem{Rogers:2015dla}
	Adam Rogers.
	\newblock {Frequency-dependent effects of gravitational lensing within plasma}.
	\newblock {\em Mon. Not. Roy. Astron. Soc.}, 451(1):17--25, 2015.
	
	\bibitem{Rogers:2016xcc}
	Adam Rogers.
	\newblock {Escape and Trapping of Low-Frequency Gravitationally Lensed Rays by
		Compact Objects within Plasma}.
	\newblock {\em Mon. Not. Roy. Astron. Soc.}, 465(2):2151--2159, 2017.
	
	\bibitem{Rogers:2017ofq}
	Adam Rogers.
	\newblock {Gravitational Lensing of Rays through the Levitating Atmospheres of
		Compact Objects}.
	\newblock {\em Universe}, 3:3, 2017.
	
	\bibitem{2018MNRAS.475..867E}
	X.~{Er} and A.~{Rogers}.
	\newblock {Two families of astrophysical diverging lens models}.
	\newblock {\em Mon.Not.Roy.Astron.Soc.}, 475:867--878, March 2018.
	
	\bibitem{Er:2013efa}
	Xinzhong Er and Shude Mao.
	\newblock {Effects of plasma on gravitational lensing}.
	\newblock {\em Mon. Not. Roy. Astron. Soc.}, 437(3):2180--2186, 2014.
	
	\bibitem{Yan:2019etp}
	Haopeng Yan.
	\newblock {Influence of a plasma on the observational signature of a high-spin
		Kerr black hole}.
	\newblock {\em Phys. Rev.}, D99(8):084050, 2019.
	
	\bibitem{bao-Finsler}
	Bao~D. Chern~S.S., Shen~Z.
	\newblock {\em An Introduction to Riemann-Finsler Geometry}.
	\newblock Graduate Texts in Mathematics 200. Springer-Verlag New York, 1
	edition, 2000.
	
	\bibitem{Gibbons-Jac-Maup-3}
	Sumanto Chanda, G.~W. Gibbons, Partha Guha, Paolo Maraner, and Marcus~C.
	Werner.
	\newblock {Jacobi-Maupertuis Randers-Finsler metric for curved spaces and the
		gravitational magnetoelectric effect}.
	\newblock Arxiv: 1903.11805 (2019).
	
	\bibitem{nazim}
	Terzioglu Nazim.
	\newblock {Uber Finslersche Raumee}.
	\newblock Wolf, Munchen (1936).
	
	\bibitem{Jusufi:2019rcw}
	Kimet Jusufi.
	\newblock {Deflection angle of charged massive particles in slowly rotating
		Kerr-Newman space-times via Gauss-Bonnet theorem and Hamilton-Jacobi method}.
	\newblock Arxiv: 1906.12186 (2019).
	
	\bibitem{Jusufi:2018kry}
	Kimet Jusufi.
	\newblock {Gravitational deflection of relativistic massive particles by Kerr
		black holes and Teo wormholes viewed as a topological effect}.
	\newblock {\em Phys. Rev.}, D98(6):064017, 2018.
	
	\bibitem{Edery2006}
	A.~Edery and J.~Godin.
	\newblock Second order kerr deflection.
	\newblock {\em General Relativity and Gravitation}, 38(11):1715--1722, Nov
	2006.
	
	\bibitem{doi:10.1063/1.1705193}
	Robert~H. Boyer and Richard~W. Lindquist.
	\newblock Maximal analytic extension of the kerr metric.
	\newblock {\em Journal of Mathematical Physics}, 8(2):265--281, 1967.
	
	\bibitem{Aazami:2011tw}
	Amir~B. Aazami, Charles~R. Keeton, and A.~O. Petters.
	\newblock {Lensing by Kerr Black Holes. II: Analytical Study of
		Quasi-Equatorial Lensing Observables}.
	\newblock {\em J. Math. Phys.}, 52:102501, 2011.
	
	\bibitem{Ono:2018jrv}
	Toshiaki Ono, Asahi Ishihara, and Hideki Asada.
	\newblock {Deflection angle of light for an observer and source at finite
		distance from a rotating global monopole}.
	\newblock {\em Phys. Rev.}, D99(12):124030, 2019.
	
	\bibitem{Jusufi:2017mav}
	Kimet Jusufi and Ali {\"O}vg{\"u}n.
	\newblock {Gravitational Lensing by Rotating Wormholes}.
	\newblock {\em Phys. Rev.}, D97(2):024042, 2018.
	
	\bibitem{Iyer:2009hq}
	Savitri~V. Iyer and Edward~C. Hansen.
	\newblock {Strong and Weak Deflection of Light in the Equatorial Plane of a
		Kerr Black Hole}.
	\newblock Arxiv: 0908.0085 (2009).
	
	\bibitem{Crisnejo-finite}
	Gabriel Crisnejo, Emanuel Gallo, and Adam Rogers.
	\newblock Finite distance corrections to the light deflection in a
	gravitational field with a plasma medium.
	\newblock {\em Phys. Rev. D}, 99:124001, Jun 2019.
	
	\bibitem{Psaltis_2012}
	Dimitrios Psaltis.
	\newblock The influence of gas dynamics on measuring the properties of the
	black hole in the center of the milky way with stellar orbits and pulsars.
	\newblock {\em The Astrophysical Journal}, 759(2):130, Oct 2012.
	
	\bibitem{10.1093/mnras/stz138}
	Tom Kimpson, Kinwah Wu, and Silvia Zane.
	\newblock {Spatial dispersion of light rays propagating through a plasma in
		Kerr space–time}.
	\newblock {\em Monthly Notices of the Royal Astronomical Society},
	484(2):2411--2419, 01 2019.
	
	\bibitem{Tyler-1977}
	G.~Leonard Tyler, Joseph~P. Brenkle, Thomas~A. Komarek, and Arthur~I.
	Zygielbaum.
	\newblock {The Viking Solar Corona Experiment}.
	\newblock {\em Journal of Geophysical Research}, 82:4335--4340, 1977.
	
	\bibitem{giampieri1996}
	G.~{Giampieri}.
	\newblock {Relativity Experiments in the Solar System}.
	\newblock In M.~{Carfora}, M.~{Cavagli{\`a}}, P.~{Fr{\'e}}, G.~{Pizzella},
	C.~{Reina}, and A.~{Treves}, editors, {\em General Relativity and
		Gravitational Physics}, page 181, 1996.
	
	\bibitem{Bertotti-1998}
	Bruno Bertotti;~Giacomo Giampieri.
	\newblock Solar coronal plasma in doppler measurements.
	\newblock {\em Solar Physics}, 178, 02 1998.
	
	\bibitem{Turyshev_2019}
	Slava~G Turyshev and Viktor~T Toth.
	\newblock Diffraction of light by plasma in the solar system.
	\newblock {\em Journal of Optics}, 21(4):045601,  2019.
	
\end{thebibliography}

\end{document}